\documentclass[prl,reprint,superscriptaddress]{revtex4-2}

\usepackage[utf8]{inputenc}
\usepackage{graphicx}

\graphicspath{{./figures/}}

\usepackage{amsmath}

\usepackage[dvipsnames]{xcolor}
\definecolor{dark-blue}{rgb}{0,0.2,0.6}

\usepackage[colorlinks,%
            pdfusetitle,%
            urlcolor=dark-blue,%
            citecolor=dark-blue,%
            linkcolor=dark-blue]{hyperref}

\usepackage{etoolbox}
\makeatletter
\pretocmd{\NAT@open}{\begingroup\color{\@citecolor}}{}{}
\apptocmd{\NAT@close}{\endgroup}{}{}
\makeatother

\makeatletter
\c@secnumdepth=+\maxdimen%
\makeatother

\makeatletter
\@booleantrue\acknowledgments@sw%
\makeatother

\newcommand{\ket}[1]{\ensuremath{\left|{#1}\right\rangle}}

\newcommand{\yb}{\ensuremath{{^\text{171}\text{Yb}}}}

\newcommand{\tP}[1]{\ensuremath{{^3\text{P}_{#1}}}}
\newcommand{\sS}[1]{\ensuremath{{^1\text{S}_{#1}}}}

\newcommand{\asciimathunit}[1]{\ensuremath{\,\text{#1}}}

\newcommand{\nm}{\asciimathunit{nm}}

\newcommand{\Hz}{\asciimathunit{Hz}}
\newcommand{\kHz}{\asciimathunit{kHz}}
\newcommand{\Gauss}{\asciimathunit{G}}

\newcommand{\Erec}{E_\text{rec}}

\newcommand{\subfigref}[2]{\hyperref[fig:#1]{\ref*{fig:#1}(#2)}}

\begin{document}


\title{Probing transport and slow relaxation in the mass-imbalanced Fermi-Hubbard model}

\author{N.~\surname{Darkwah Oppong}}\email{n.darkwahoppong@lmu.de}
\author{G.~Pasqualetti}
\author{O.~Bettermann}
\affiliation{Ludwig-Maximilians-Universit{\"a}t, Schellingstra{\ss}e 4, 80799 M{\"u}nchen, Germany}
\affiliation{Max-Planck-Institut f{\"u}r Quantenoptik, Hans-Kopfermann-Stra{\ss}e 1, 85748 Garching, Germany}
\affiliation{Munich Center for Quantum Science and Technology (MCQST), Schellingstra{\ss}e 4, 80799 M{\"u}nchen, Germany}

\author{P.~Zechmann}
\author{M.~Knap}
\affiliation{Department of Physics and Institute for Advanced Study, Technical University of Munich, 85748 Garching, Germany}
\affiliation{Munich Center for Quantum Science and Technology (MCQST), Schellingstra{\ss}e 4, 80799 M{\"u}nchen, Germany}

\author{I.~Bloch}
\author{S.~F{\"o}lling}
\affiliation{Ludwig-Maximilians-Universit{\"a}t, Schellingstra{\ss}e 4, 80799 M{\"u}nchen, Germany}
\affiliation{Max-Planck-Institut f{\"u}r Quantenoptik, Hans-Kopfermann-Stra{\ss}e 1, 85748 Garching, Germany}
\affiliation{Munich Center for Quantum Science and Technology (MCQST), Schellingstra{\ss}e 4, 80799 M{\"u}nchen, Germany}

\hypersetup{pdfauthor={N.~Darkwah~Oppong, G.~Pasqualetti, O.~Bettermann, P.~Zechmann, M.~Knap, I.~Bloch, S.~Fölling}}

\date{\today}

\begin{abstract}
Constraints in the dynamics of quantum many-body systems can dramatically alter transport properties and relaxation timescales even in the absence of static disorder.
Here, we report on the observation of such constrained dynamics arising from the distinct mobility of two species in the one-dimensional mass-imbalanced Fermi-Hubbard model, realized with ultracold ytterbium atoms in a state-dependent optical lattice.
By displacing the trap potential and monitoring the subsequent dynamical response of the system, we identify suppressed transport and slow relaxation with a strong dependence on the mass imbalance and interspecies interaction strength, consistent with eventual thermalization for long times.
Our observations demonstrate the potential for quantum simulators to provide insights into unconventional relaxation dynamics arising from constraints.
\end{abstract}

\maketitle


\section{Introduction}
The fundamental understanding of thermalization and its failure in isolated quantum many-body systems has seen remarkable progress in recent years, driven both by novel theoretical concepts~\cite{deutsch:1991, srednicki:1994, rigol:2008, dalessio:2016} and emerging experimental platforms for quantum simulations such as ultracold atoms, trapped ions, and superconducting qubits~\cite{bloch:2008, blatt:2012, kjaergaard:2020}.
The simplest approach to the problem is to posit the two distinct classes of ergodic and nonergodic dynamics.
In the former case, fast relaxation to a local equilibrium is followed by slow thermalization of globally conserved quantities according to the laws of hydrodynamics~\cite{hohenberg:1977, mukerjee:2006, lux:2014, bohrdt:2017}.
By contrast, nonergodic dynamics can arise in systems with a large number of conserved quantities, as exemplified by many-body localization in strongly disordered quantum systems~\cite{schreiber:2015, kondov:2015, smith:2016, roushan:2017} that retain their memory of an initial state for arbitrarily long times~\cite{nandkishore:2015, abanin:2019, serbyn:2014}.

\begin{figure}[b] 
	\includegraphics[width=\columnwidth]{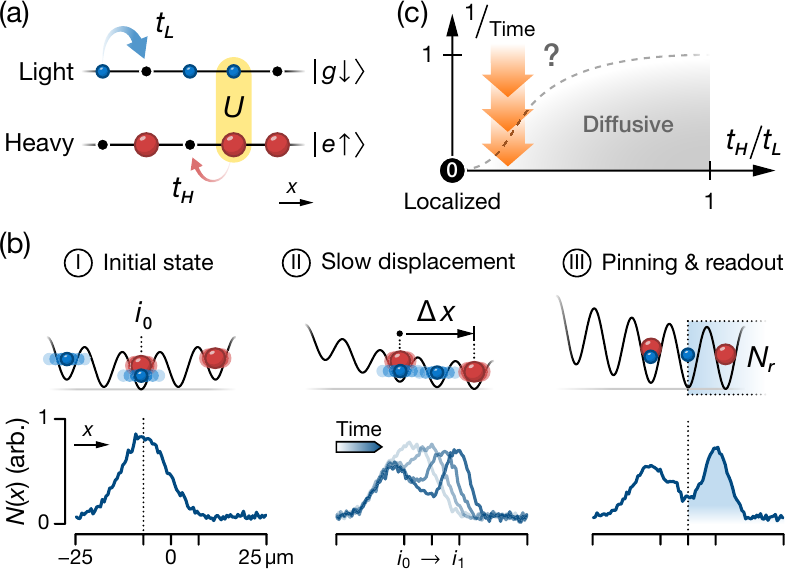}
	\caption{\label{fig:schematic}%
    Transport in the mass-imbalanced Fermi-Hubbard model.
	(a)~Illustration of the mass-imbalanced Fermi-Hubbard model with on-site interaction $U$ and hopping \mbox{$t_L \gg t_H$} of the light (blue circles) and heavy (red circles) particles, respectively.
    We also show the atomic states relevant for the experimental implementation.
	(b)~Schematic representation of the different steps in our transport measurement (top row) and integrated densities of the light atoms (bottom row) from experimental snapshots.
	(c)~Scenario for the transient regime following a quench in the mass-imbalanced Fermi-Hubbard model for variable inverse hold time (vertical axis).
	Here, the dashed line marks the crossover between unconventional relaxation dynamics at early times (question mark) and diffusion at late times.
	The orange arrows indicate a trajectory in the regime of our experiment.}
\end{figure}

Recent experimental and theoretical investigations of nonequilibrium dynamics in isolated quantum many-body systems, however, suggest considerable refinements to this classification.
For instance, Rydberg quantum simulators have observed surprising oscillatory dynamics in the blockade regime~\cite{bernien:2017, turner:2018}, and the slow late-time dynamics of fractonic quantum matter with constrained excitations~\cite{nandkishore:2019} is described by substantially modified hydrodynamic equations~\cite{gromov:2020, feldmeier:2020a, guardado-sanchez:2020}.
Furthermore, interacting mixtures of heavy and light particles [see Fig.~\subfigref{schematic}{a}] have even been proposed to feature a dynamical type of many-body localization arising from the heavy particles acting as a form of disorder for the light ones~\cite{kagan:1984, de-roeck:2014, schiulaz:2014, schiulaz:2015}.
Although subsequent studies have suggested that such heavy-light mixtures are ergodic and do thermalize at late times, the relaxation of initial nonequilibrium states is expected to be unconventionally slow~\cite{de-roeck:2014a, schiulaz:2014, papic:2015, yao:2016, sirker:2019}.
This is a direct consequence of the strongly constrained motion in the presence of mass imbalance and interactions.
However, since such heavy-light mixtures are particularly challenging to simulate with classical resources, these theoretical studies do not entirely agree on the exact properties of relaxation~\cite{papic:2015, yao:2016, sirker:2019}, which underlines the importance of experimental evidence. 
Crucially, this phenomenology connects to the general question of how dynamical constraints can introduce slow equilibration and nonergodicity in quantum many-body systems~\mbox{\cite{lan:2018, feldmeier:2019, pancotti:2020, guardado-sanchez:2021, scherg:2021, morong:2021}}.

In this work, we experimentally study the nonequilibrium dynamics of a heavy-light mixture with an ultracold quantum gas of \yb{} atoms in an optical lattice.
Our experiment realizes the strongly mass-imbalanced Fermi-Hubbard model in one dimension~(1D), as illustrated in Fig.~\subfigref{schematic}{a}.
In addition to the \sS0 ground state (denoted $\ket{g}$), the alkaline-earth-like atom \yb{} features the metastable \tP0 excited state (denoted $\ket{e}$), often referred to as the clock state~\cite{ludlow:2015}.
We harness a state-dependent optical lattice (SDL)~\cite{riegger:2018} to introduce different timescales for the hopping of atoms in $\ket{g}$ and $\ket{e}$ taking on the role of light and heavy particles, respectively.
Our system is harmonically confined, and we gradually displace the trap minimum with the help of an additional optical potential to probe the transport dynamics of the light species [see Fig.~\subfigref{schematic}{b}].
Measuring the dynamics of the atomic cloud, we find  relaxation at late times as a signature of ergodicity for all finite interaction parameters.
However, this relaxation becomes slow compared to conventional timescales for strong interactions and large mass imbalance.
Our observations cannot be explained by a mere reduction of the mobility for the heavy species alone, but rather they can be understood as emergent properties from the constrained many-body dynamics.
Numerical simulations based on matrix-product states and exact diagonalization support our findings.

\section{Experiment}
Our experiment begins with a Fermi gas of \yb{} atoms in a balanced mixture of the two nuclear spins $\ket{m_F=-1/2} \equiv \ket{\downarrow}$ and $\ket{m_F=+1/2} \equiv \ket{\uparrow}$ in the ground state $\ket{g}$.
In the initial optical dipole trap, we prepare a total of approximately $10^4$ atoms at a temperature of $T \simeq 0.15 T_F$, with $T_F$ the Fermi temperature.
The weakly interacting spin mixture is loaded from the optical dipole trap into the ground band of a two-axes, approximately $30\Erec^m$ deep, state-independent lattice operated at the magic wavelength ($\lambda_m = 759.3\nm$)~\cite{barber:2008} and a $V_L=6.9(3)\Erec$ deep SDL ($\lambda = 671.5\nm$) along the third axis.
Here, $\Erec^m = h \times 2.0\kHz$ and $\Erec = h \times 2.6\kHz$ are the recoil energies of the corresponding lattice photons.
The atoms are nonuniformly distributed across the array of decoupled 1D systems (tubes) generated by the perpendicular state-independent lattices.
We estimate approximately $300$ tubes are considerably filled with a mean atom number of $\mathcal{N} \simeq 18$ per spin state and standard deviation $\Delta\mathcal{N} \simeq 8$.
In a typical tube, the atoms are spread over a system size of $l \approx 30$ lattice sites (root-mean-square width)~\cite{SM}.

Shortly after loading the lattices, we use a $0.17$-ms-long clock laser pulse to selectively drive atoms from $\ket{g \uparrow}$ to $\ket{e \uparrow}$.
This suddenly introduces a distinct hopping amplitude $t_H \simeq h\times 5\Hz \ll t_L$ for the clock state atoms in $\ket{e\uparrow} \equiv \ket{H}$ (heavy).
Here, $t_L \simeq h\times 105\Hz$ is the unaltered hopping amplitude of the remaining ground-state atoms in $\ket{g \downarrow} \equiv \ket{L}$ (light).
The ratio of the hopping amplitudes and the associated mass imbalance is determined by the different lattice depths $V_L$ and $V_H = 3.06(4) V_L$ experienced by $\ket{L}$ and $\ket{H}$ atoms in the SDL~\cite{SM}.
We ensure that the mixture initially remains noninteracting by ramping the magnetic field to the zero crossing of an orbital Feshbach resonance at approximately~$1533\Gauss$~\cite{bettermann:2020,SM} before the excitation pulse.
Finally, we lower the SDL depth to adjust the hopping ratio $t_H/t_L$ and slowly ramp the magnetic field to $1400$--$1600\Gauss$ in accordance with the desired interaction strength.
We note that the fraction of doublons as a critical property of the initial state only weakly depends on the chosen interaction parameter~\cite{SM}.
After the state preparation procedure, we translate the minimum of the trapping potential along the tubes by slowly turning on a displaced, state-independent dipole trap, which initiates the transport dynamics.
Following the subsequent evolution, we rapidly ramp up the SDL to $V_L \approx 15\Erec$, which freezes the motion of light and heavy atoms.
We then turn off the magnetic field and record the density of the light atoms with \textit{in situ} absorption imaging and determine the fraction transported to the new trap minimum [see Fig.~\subfigref{schematic}{b}].
Our imaging intrinsically integrates along one axis of the system, which averages the measurement over an ensemble of tubes with different atom numbers~\cite{SM}.

\section{Model}
Each tube in our experiment is described by a 1D Fermi-Hubbard model with mass imbalance,
\begin{align}\label{eq:hamiltonian}
  \begin{split}
  \hat{\mathcal{H}} &=
    -\sum_{i,\alpha\in\{L,H\}}t_\alpha \left[\hat{c}_{i\alpha}^\dagger \hat{c}^{\phantom\dagger}_{(i+1)\alpha} + \text{H.c.}\right] \\
    &\quad\,+ U \sum_i \hat{n}_{iL} \hat{n}_{iH} 
    + \frac{\kappa}{2} \sum_{i,\alpha\in\{L,H\}} {(i-i_0)}^2\hat{n}_{i\alpha}.
  \end{split}
\end{align}
Here, $\hat{c}^{\phantom\dagger}_{i\alpha}$ ($\hat{c}^\dagger_{i\alpha}$) denotes the fermionic creation (annihilation) operator for a light ($\alpha=L$) or heavy ($\alpha=H$) atom, $\hat{n}_{i\alpha} \equiv \hat{c}^\dagger_{i\alpha} \hat{c}^{\phantom\dagger}_{i\alpha}$ the corresponding number operator, $t_\alpha$ the hopping amplitude of each species, and $U$ the on-site interaction.
The harmonic confinement is determined by the trap minimum $i_0$ and strength $\kappa=m \omega^2 d^2 = h\times 3.1(1)\Hz$, with atomic mass $m$, trapping frequency $\omega = 2\pi \times 40(1)\Hz$, and lattice spacing $d = \lambda / 2$.

For a completely frozen heavy species ($t_H = 0$), the Hamiltonian~$\hat{\mathcal{H}}$ decouples into single-particle terms for the light species.
In this limit, known as the Falicov-Kimball model for $\kappa=0$~\cite{falicov:1969}, the heavy particles only contribute as an effective binary disorder potential giving rise to single-particle localization~\cite{heitmann:2020}.
Previous experiments have employed binary mixtures of bosons to probe such a regime, with disorder formed by a fully localized species~\cite{gadway:2011}.
In contrast, we study a system of two mobile species with distinct hopping amplitudes~$t_L \gg t_H > 0$ [see Fig.~\subfigref{schematic}{c}], realizing two strongly differing but finite scales in the Hamiltonian.
For finite interaction strength~$|U| > 0$, numerical simulations have suggested that the resulting dynamical constraints can lead to anomalous transport and unconventionally slow thermalization with a nontrivial dependence on the parameters of the model~\cite{de-roeck:2014a, schiulaz:2014, papic:2015, yao:2016, sirker:2019}.
At late times, this behavior is expected to be superseded by diffusive transport, as shown in Fig.~\subfigref{schematic}{c}.
Interestingly, for large interactions and strongly differing masses, the crossover timescale is expected to become extraordinarily large---a genuine many-body effect~\cite{SM}.
In the following measurements, we probe the nature of this transient regime at different times with parameters $t_H/t_L \gtrsim 0.1$ and $|U|/t_L \lesssim 10$.

\section{Localized single-particle states}
In a first reference measurement, we characterize the single-particle physics originating from the harmonic confinement $\kappa > 0$ in our experiment. It gives rise to single-particle localized eigenstates at the edge of the trap, a phenomenon studied in Refs.~\cite{pezze:2004, ott:2004, orso:2004, rigol:2004, rey:2005, schulz:2016}.
These Wannier-Stark-localized states occur because of the finite gradient $\partial_i \hat{\mathcal{H}} = \kappa (i - i_0) \hat{n}_{i\alpha}$ between neighboring lattice sites~\cite{wannier:1962}.
To experimentally probe this effect, we prepare a clean sample of noninteracting light atoms by employing a short resonant ``push'' pulse that removes all $\ket{g\uparrow}$ atoms.
Subsequently, we measure the response to a linear translation of the trap minimum as shown in Fig.~\subfigref{schematic}{b}.
Here, the trap minimum is displaced by $\Delta x/d = i_1-i_0= 47(3)$ within time $90(5)\hbar/t_L$ and we consider variable $\kappa/t_L$ by adjusting the SDL depth in the range $0.69(3)$--$11.8(5)\Erec$.

\begin{figure}[t!]
  \includegraphics[width=\columnwidth]{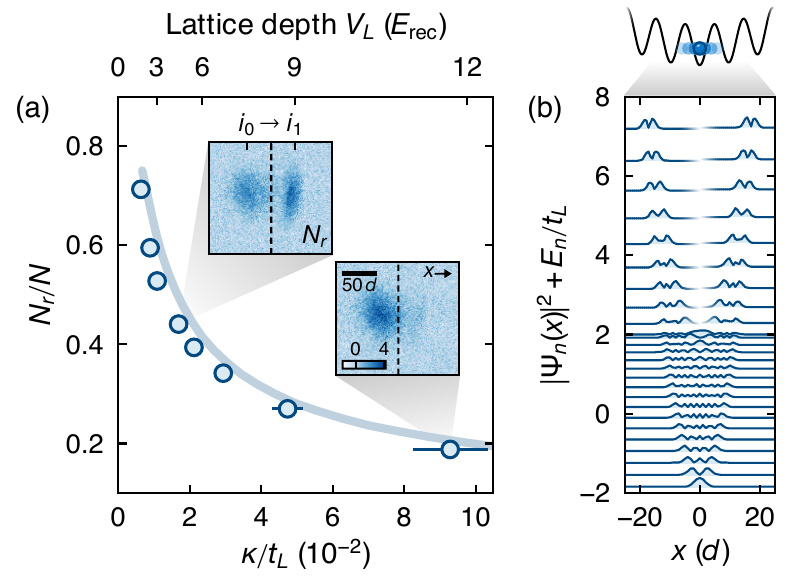}
  \caption{\label{fig:non-int}%
    Wannier-Stark localization in the absence of interactions.
    (a)~Fraction of light atoms $N_r/N$ transported to the right half of the system (blue circles) in the absence of heavy atoms and for variable confinement $\kappa/t_L$ determined by the lattice depth~$V_L$.
    Each point is the average of $2$--$5$ measurements, and error bars indicate the uncertainty of $\kappa/t_L$ (partly smaller than the marker size).
    The solid line corresponds to our numerical calculation~\cite{SM}, and the insets show raw atomic column densities for $\kappa/t_L = 1.7(1) \times 10^{-2}$ and $9(1) \times 10^{-2}$ with the dashed lines indicating the boundary that determines the atom count $N_r$ in the right half of the system.
    (b)~Probability density ${|\Psi_n(x)|}^2$ and eigenenergy $E_n/t_L$ of the lowest-lying single-particle eigenstates for $V_L = 9\Erec$ ($\kappa/t_L = 4.9 \times 10^{-2}$).
    We add a small linear potential to lift the degeneracy of the localized states.
  }
\end{figure}

For all finite lattice depths, we observe a separation of the atoms into two clouds, one close to the initial trap minimum $i_0$ and one at the final trap minimum $i_1$, as visible in the insets of Fig.~\subfigref{non-int}{a}.
We quantify this by determining the number of atoms in the right half of the system $N_r$, which corresponds to $i \geq i_0 + \Delta x/(2d)$, and compare it to the total number of atoms $N$.
When measuring $N_r$ before the displacement of the trap minimum, we find a nearly constant value of approximately $10\%$, which we attribute to our finite imaging resolution and a small extent of the initial cloud into the counting region.
In Fig.~\subfigref{non-int}{a}, we plot the fraction of transported atoms, $N_r/N$, which is exponentially suppressed for increasing confinement strength.
This significant reduction of mass transport in the system results from the properties of the single-particle eigenstates, which are visualized in Fig.~\subfigref{non-int}{b}.
In general, eigenstates with energies $E_n \in [-2t_L, 2t_L]$ are delocalized across the trap with a nonzero probability density at the center.
This behavior changes dramatically for energies $E_n > 2t_L$ corresponding to localized states, in which atoms cannot efficiently follow the displaced trap minimum.
Crucially, the number of states with $E_n \leq 2t_L$ is reduced as we increase $\kappa/t_L$ and therefore, a larger number of localized states becomes occupied in each tube.
This explains the observed suppression of $N_r/N$, which we also reproduce with a theoretical calculation~\cite{SM} of the experimental protocol~[see Fig.~\subfigref{non-int}{a}].

\section{Inhibited transport at early times}
Next, we probe how interactions with the heavy species modify the mobility of the light atoms.
For this measurement, we set the interaction strength $U/t_L \in [-20, 5]$, a range accessible via magnetic-field tuning of the orbital Feshbach resonance, and choose the fixed hopping ratio $t_H/t_L \simeq 0.1$ and confinement $\kappa/t_L \simeq 1.7 \times 10^{-2}$ by ramping the SDL to $4.7(2)\Erec$.
Here, the confinement~$\kappa/t_H \simeq 10 \kappa/t_L = 1.7 \times 10^{-1}$ suggests that most heavy atoms occupy single-particle localized states according to our previous measurement in the noninteracting regime [see Fig.~\subfigref{non-int}{a}].
However, in the presence of interactions, several of these states delocalize in analogy to the suggested critical gradient required for localization of the interacting Wannier-Stark ladder~\cite{ott:2004, strohmaier:2007, schulz:2019, van_nieuwenburg:2019, chanda:2020}.
We numerically verify this for our mass-imbalanced case and finite but weak interactions.
Here, we find a central trap region of approximately $30$ lattice sites, in which the heavy atoms already become mobile during state preparation~\cite{SM}.
Thus, a significant fraction of the system will be responsive to the displacement of the trap potential. 

\begin{figure}[t!]
  \includegraphics[width=\columnwidth]{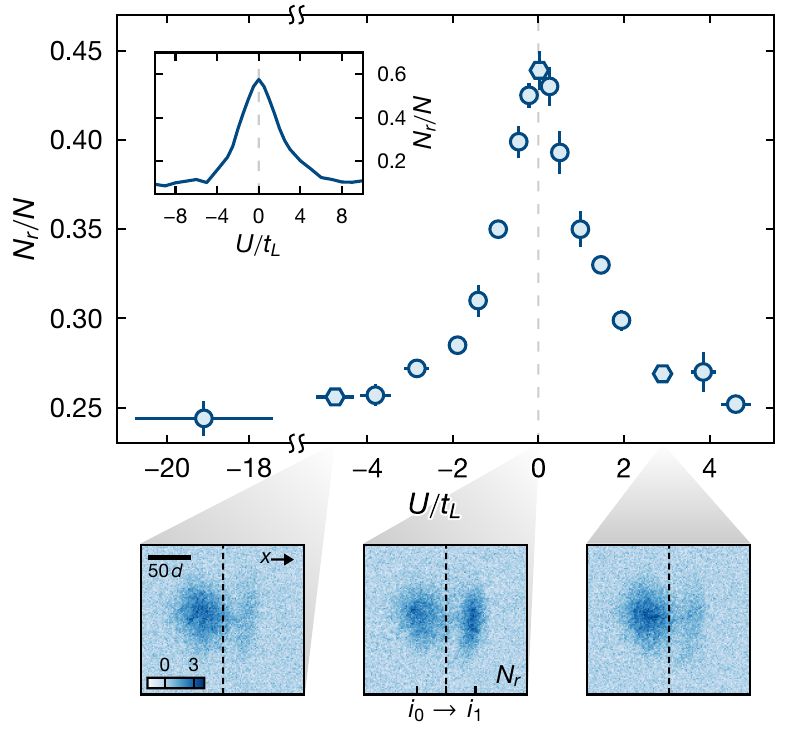}
  \caption{\label{fig:int}%
    Constrained early-time dynamics in the interacting heavy-light mixture.
    Fraction of light atoms $N_r/N$ transported to the right half of the system (blue markers) for variable heavy-light interaction strength $U$, fixed hopping ratio $t_H/t_L = 0.104(7)$, and confinement $\kappa/t_L = 1.7(1) \times 10^{-2}$.
    Each point is the average of $3$--$4$ measurements, and error bars (partly smaller than the marker size) denote the standard error of the mean of $N_r/N$ and the uncertainty of $U/t_L$.
    The bottom panels show raw atomic column densities for $U/t_L = -4.8(4)$, $0.02(4)$, and $2.9(2)$ (from left to right, hexagonal markers) with the dashed lines indicating the boundary that determines the atom count $N_r$ in the right half of the system.
    The inset of the main panel shows the numerical matrix-product-state simulations of a single tube with the Hubbard parameters and confinement strength of the experiment and $\mathcal{N}_{\rm sim}=5$ atoms of each species.
  }
\end{figure}

In analogy to our first measurement, we again determine the fraction of transported atoms after displacing the trap minimum by $47(3)d$ within time $92(5)\hbar / t_L$.
As we increase the interaction strength $|U|/t_L$, we find a substantial reduction of the transported fraction $N_r/N$ up to a factor of 2 compared to $U=0$ (see Fig.~\ref{fig:int}).
The behavior for attractive ($U < 0$) and repulsive ($U > 0$) interactions is almost identical, resulting from a similar initial state~\cite{SM} and the dynamical symmetry of our model in the limit $\kappa=0$~\cite{schneider:2012}.
We find that the most significant change in the fraction of transported atoms occurs for interaction energies $|U|/t_L < 2$, whereas the signal saturates and remains nearly constant at a low value for $|U|/t_L > 4$.
To support our experimental data, we perform matrix-product-state simulations for the Fermi-Hubbard model described in Eq.~\eqref{eq:hamiltonian} and a simplified version of the experimental protocol~\cite{SM}.
As shown in the inset of Fig.~\ref{fig:int}, the extracted $N_r/N$ curve agrees qualitatively with the experiment.
Quantitative disagreement is expected due to the significantly lower atom number in the simulation, $\mathcal{N}_{\rm sim} = 5 \ll \mathcal{N}$.

In our measurement, the apparent suppression of transport results from the constraints in the dynamics of the light atoms arising from interactions with the heavy species, as explored in Fig.~\subfigref{schematic}{c}.
We emphasize that this measurement primarily probes the early-time dynamics on the timescale $\hbar/t_H$ since the system is effectively frozen and fully separated towards the end of the gradual displacement over the relatively large distance $\Delta x/d \simeq 47$, which exceeds the typical system size $l \approx 30$ (see bottom panels of Fig.~\ref{fig:int}).

\section{Slow relaxation at late times}
We now turn to the dynamical response of the heavy-light mixture and investigate whether light atoms, remaining close to their initial position during the trap translation, relax towards the final trap minimum at much later times.
Such dynamics lead to signatures in the \textit{in situ} density, which we systematically record for a variable hold time.
Here, we reduce the displacement to $\Delta x/d$ $\simeq$ $ 20 < l$ while keeping the speed approximately $0.5d (t_L/\hbar)$ unchanged to ensure that the system remains connected and a significant fraction of the heavy atoms are mobile over the traversed distance.
The relaxation dynamics can be captured by quantifying the change of the \textit{in situ} density distribution between the initial state after the translation of the trap minimum at a hold time $\tau = 0$ and later times $0 < \tau \lesssim 400\hbar/t_L$.
To this end, we introduce the density deviation observable,
\begin{align}\label{eq:dens-dev}
  \delta n{(\tau)} = {\left\{ {\int dx\,n(x, \tau) {[ n(x, \tau) - n(x, 0) ]}^2} \right\}}^{(1/2)},
\end{align}
which is obtained from the density of the light atoms integrated perpendicularly to the transport direction $x$ and normalized such that \mbox{$\int dx\,n(x, \tau) = 1$}.
For a slow relaxation of the density, this observable grows monotonously with increasing $\tau$, whereas a constant $\delta n(\tau)$ indicates a stationary state of the system.

\begin{figure*}[t]
  \includegraphics[width=\textwidth]{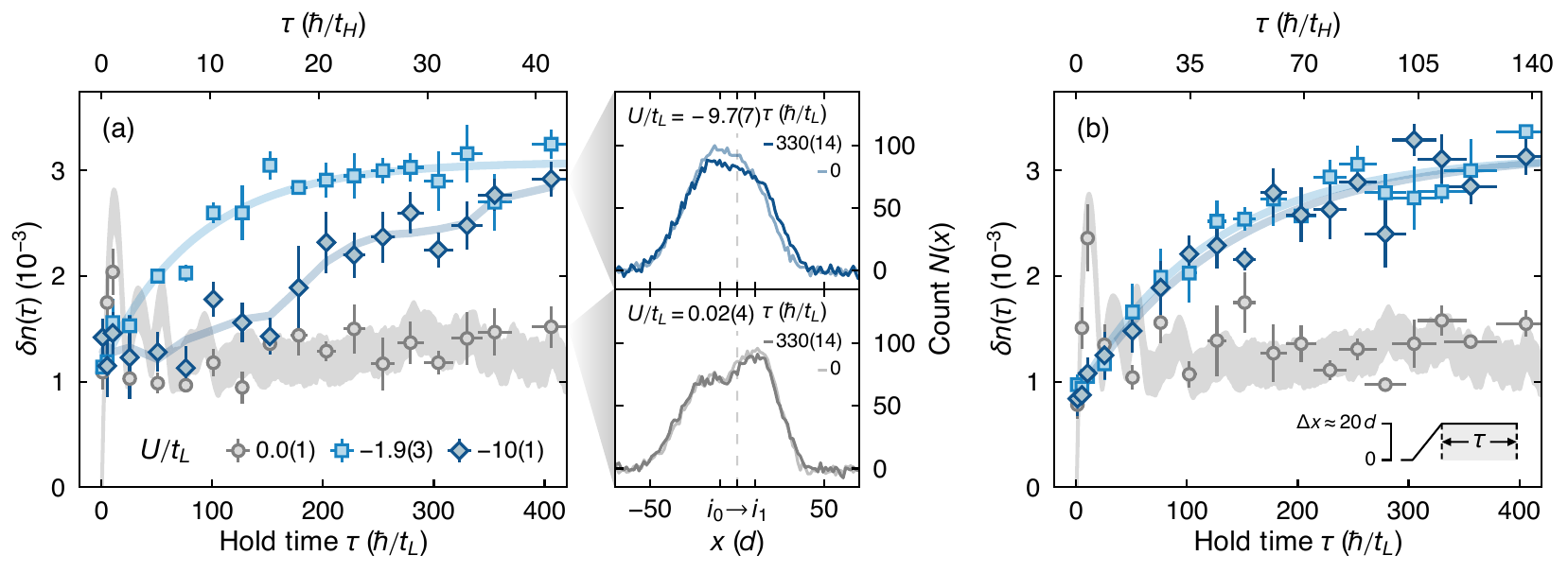}
  \caption{\label{fig:dynamics}%
    Late-time relaxation dynamics.
    Density deviation $\delta n (\tau)$ of the light atoms with respect to the initial atom distribution after translating the trap minimum by $\Delta x$ [see Eq.~\eqref{eq:dens-dev}], for a variable hold time $\tau > 0$, as illustrated in the bottom-right schematic of panel~(b).
    The time traces correspond to three heavy-light interaction strengths (see legend) with (a)~$t_H/t_L = 0.102(6)$ and (b)~$0.34(3)$.
    Each data point is calculated from the average of $3$-$5$ atomic densities and an independent measurement of $n(x, 0)$.
    Error bars (partly smaller than the marker size) denote the uncertainty of $\tau$, determined from the estimated uncertainty of the hopping amplitude~$t_L$, and standard error of $\delta n (\tau)$, determined from jackknife resampling~\cite{SM}.
    The colored lines show exponential fits, as discussed in the main text, and in panel~(a), a three-point moving average for $U/t_L\simeq-10$ as a guide to the eye.
    We show our numerical calculation for $U=0$ as a gray band with the height derived from the estimated systematic uncertainty of $\delta n (\tau)$~\cite{SM}.
    The two central panels show integrated atomic densities, which are used to calculate $\delta n(\tau=330 \hbar/t_L)$ in~(a).
  }
\end{figure*}

We probe the influence of the interaction strength on the dynamics by measuring the density deviation for $t_H/t_L \simeq 0.1$ and $U/t_L\simeq 0$, $-2$, $-10$, as shown in Fig.~\subfigref{dynamics}{a}.
First, we focus on the noninteracting time trace with $U/t_L\simeq0$, which shows transient dynamics of $\delta n(\tau)$ at early times for both hopping ratios.
These large-amplitude transients are quickly damped due to contributions from different eigenstates.
For later times $\tau \gtrsim 100\hbar/t_L$, the density deviation reaches an almost stationary signal, which we also expect from our numerical calculation.

For finite interaction strengths, we find a contrasting behavior with a monotonously increasing density deviation~$\delta n(\tau)$ for $\tau > 0$.
In this regime, the integrated atomic densities $n(x, \tau)$ show that the light atoms drift towards the final trap minimum at late times [see panels right of Fig.~\subfigref{dynamics}{a}].
To extract the timescale of the associated relaxation dynamics, we fit the experimental data to the function $\delta n(\tau) = a (1 - e^{-\tau/\tau_0}) + c$ with the exponential decay constant $\tau_0$ and parameters~$a$,~$c$.
For $U/t_L\simeq-2$, we find good agreement between the fitted function and the experimental data with $\tau_0=85(17)\hbar/t_L$.
In contrast, for the larger interaction strength~$U/t_L\simeq-10$, the fit parameters~$a$ and~$\tau_0$ take unreasonably large values yielding a linear curve.
Until~$\tau\approx100\hbar/t_L$, the measured density deviation grows far less compared to the case with~$U/t_L\simeq-2$.
To compare the dynamics for the two interaction strengths, we fit the measured density deviation to a linear curve for early times $\tau \leq 101\hbar/t_L$~\cite{SM}.
Here, we find the linear slope $2(2) \times 10^{-6} \,t_L/\hbar$ for $U/t_L \simeq -10$ compared to $13(2) \times 10^{-6}\, t_L/\hbar$ for $U/t_L \simeq -2$.
The significantly smaller linear slope illustrates the emergence of unconventionally slow dynamics for strong interactions.

To demonstrate that the slow relaxation stems from dynamical constraints induced by mass imbalance, we also measure the density deviation for the significantly larger~$t_H/t_L= 0.34(3)$.
For this value of the hopping ratio, the transient regime of unconventional relaxation is expected to be significantly shorter than for the larger ratio studied previously, as illustrated in Fig.~\subfigref{schematic}{c}.
By changing the wavelength of the SDL to $\lambda^\prime = 690.1\nm$ such that $V_H = 1.97(5)V_L$, we realize parameters closely matching our first measurement apart from the hopping ratio~$t_H/t_L$~\cite{SM}.
Because of this, the confinement strength $\kappa/t_L\simeq 1.7\times 10^{-2}$ also remains unchanged, ensuring directly comparable systems, in particular, regarding Wannier-Stark-localized states of light atoms.
As shown in Fig.~\subfigref{dynamics}{b}, for $U/t_L\simeq0$ and $-2$, the time-dependent behavior remains largely similar to the previous measurement.
However, the time trace for $U/t_L \simeq -10$ now only differs negligibly from the one for weaker interactions.
From the numerical fit of $\delta n (\tau)$, we extract similar decay constants $\tau_0 = 120(22)\hbar/t_L$ and~$138(26)\hbar/t_L$ for the two interaction strengths $U/t_L=-2$ and~$-10$, agreeing with our qualitative observation.
The experimental data for $t_H/t_L \simeq 0.3$ confirm a strong dependence of the relaxation dynamics on the interaction strength and mass imbalance between the heavy and light particles.
Hence, the unconventionally slow transport emerging in our system can be ascribed to genuine many-body effects.

This observation is further supported by considering the limit of strong interactions and mass imbalances much larger than probed experimentally.
In this regime, small-scale numerical calculations predict a pronounced change in the relaxation dynamics of a density modulation, setting in at the timescale~$\overline{\tau} \sim \hbar U / t_H^2$ (valid for $U \gg t_L$)~\cite{yao:2016}.
This timescale arises from the collective motion of a bound state comprised of two heavy particles on neighboring sites and a light one delocalized over these sites~\cite{schiulaz:2014}.
However, we find numerical evidence that a noticeable change of dynamics at $\overline{\tau}$ vanishes with increasing system size for moderate hopping ratios~$t_H/t_L \geq 0.01$~\cite{SM}.
These findings are consistent with the experimentally observed relaxation, which does not exhibit prominent features at $\overline{\tau}$.
The smooth relaxation dynamics in larger systems could potentially originate from an increased relevance of the many-body medium and a hierarchy of relaxation timescales.
Such a hierarchy would emerge from $N$-body bound states consisting of $(N-1)$ heavy particles and one light particle, only becoming mobile at the timescale~$\hbar Ut_L^{(N-3)}/t_H^{(N-1)}$.

We briefly comment on how experimental imperfections could also drive the relaxation of $\delta n(\tau)$ in addition to the closed system dynamics of Eq.~\eqref{eq:hamiltonian}.
Most notably, we expect the mobility of the light species to increase from the small loss of heavy atoms, which becomes appreciable for $\tau \gg 100\hbar/t_L$ and reaches up to approximately $20\%$ at $\tau \simeq 400\hbar/t_L$~\cite{SM}.
However, these losses should be mostly independent of $U/t_L$ since the dominant dissipation process is the off-resonant scattering of SDL photons.
By contrast, our data show a strong interaction dependence.
Thus, we conclude that the distinct features of the time traces cannot be explained by dissipation.

\section{Discussion and outlook}
We have characterized the density dynamics of the mass-imbalanced Fermi-Hubbard model after a gradual change of the external trapping potential, which can be summarized as follows.
First, our data show relaxation of the density for all finite interaction parameters and hopping ratios at late times.
Second, for the strongest interaction $U/t_L \simeq -10$ and smallest hopping ratio $t_H/t_L \simeq 0.1$ probed in our experiment, the system relaxes much slower as a result of the dynamical constraints induced by the heavy species.
Comparing this observation to the much faster relaxation for $U/t_L \simeq -2$ shows that the slow emergent timescale must involve many-body effects.
Thus, our experimental platform gives access to a particularly interesting class of unconventionally slow dynamics that exist in the absence of static disorder and can be controlled solely by interactions.

In future experiments, variable-wavelength density modulations could provide an ideal setting to directly study anomalous transport in this system.
Moreover, it would be particularly interesting to investigate the effect of integrability on transport in the one-dimensional mass-balanced Fermi-Hubbard model and explore the weak breaking of integrability by only slightly detuning the hopping amplitudes of the two species. 
Our work also motivates the experimental study of transport in other strongly constrained states of quantum matter, which can be found in Rydberg-blockaded systems, strongly tilted lattices, and realizations of lattice gauge theories~\cite{martinez:2016, bernien:2017, guardado-sanchez:2020, scherg:2021, yang:2020, morong:2021}.


\begin{acknowledgments}
  We acknowledge the valuable and helpful discussions with Dmitry~A.~Abanin, Johannes Feldmeier, Sarang Gopalakrishnan, Bharath Hebbe Madhusudhana, Markus M\"uller, Luis Riegger, Pablo Sala, Sebastian Scherg, and Alessandro Silva, and we are grateful to Jesper Levinsen and Meera~M.~Parish for helpful insights on the orbital Feshbach resonance in an optical lattice.
  The authors also wish to thank Alexander Impertro for technical contributions to the experiment.
  This project has received funding from the Deutsche Forschungsgemeinschaft (DFG, German Research Foundation) under Germany's Excellence Strategy--EXC-2111--390814868, DFG TRR80 and DFG Grant No.~KN1254/2-1, the European Research Council (ERC) under the European Union's Horizon 2020 research and innovation programme (Grant Agreement No.~817482 and No.~851161), and the Technical University of Munich--Institute for Advanced Study, funded by the German Excellence Initiative and the European Union FP7 under Grant Agreement No.~291763.
\end{acknowledgments}

\bibliography{references,sm}


\end{document}



\title{{\small Supplemental Material} \\Probing transport and slow relaxation in the mass-imbalanced Fermi-Hubbard model}

\author{N.~\surname{Darkwah Oppong}}\email{n.darkwahoppong@lmu.de}
\author{G.~Pasqualetti}
\author{O.~Bettermann}
\affiliation{Ludwig-Maximilians-Universit{\"a}t, Schellingstra{\ss}e 4, 80799 M{\"u}nchen, Germany}
\affiliation{Max-Planck-Institut f{\"u}r Quantenoptik, Hans-Kopfermann-Stra{\ss}e 1, 85748 Garching, Germany}
\affiliation{Munich Center for Quantum Science and Technology (MCQST), Schellingstra{\ss}e 4, 80799 M{\"u}nchen, Germany}

\author{P.~Zechmann}
\author{M.~Knap}
\affiliation{Department of Physics and Institute for Advanced Study, Technical University of Munich, 85748 Garching, Germany}
\affiliation{Munich Center for Quantum Science and Technology (MCQST), Schellingstra{\ss}e 4, 80799 M{\"u}nchen, Germany}

\author{I.~Bloch}
\author{S.~F{\"o}lling}
\affiliation{Ludwig-Maximilians-Universit{\"a}t, Schellingstra{\ss}e 4, 80799 M{\"u}nchen, Germany}
\affiliation{Max-Planck-Institut f{\"u}r Quantenoptik, Hans-Kopfermann-Stra{\ss}e 1, 85748 Garching, Germany}
\affiliation{Munich Center for Quantum Science and Technology (MCQST), Schellingstra{\ss}e 4, 80799 M{\"u}nchen, Germany}

\hypersetup{pdfauthor={N.~Darkwah~Oppong, G.~Pasqualetti, O.~Bettermann, P.~Zechmann, M.~Knap, I.~Bloch, S.~Fölling}}

\date{\today}

\maketitle

\renewcommand{\thefigure}{S\arabic{figure}}
\renewcommand{\theequation}{S.\arabic{equation}}
\renewcommand{\thesection}{S.\Roman{section}}
\renewcommand{\thetable}{S\arabic{table}}

{\small \tableofcontents}

\section{Experimental techniques}

\subsection{State preparation}\label{sec:state-prep}
The details on the preparation of degenerate gases of \yb{171} are described in Ref.~\cite{bettermann:2020} and we only briefly outline the procedure here.
Our experiment starts by loading approximately $1.4 \times 10^6$ \yb{174} and $1.0 \times 10^6$  \yb{171} atoms from a magneto-optical trap (MOT) into a crossed optical dipole trap.
We perform forced evaporation by lowering the optical dipole trap.
Towards the end of the cooling sequence, we remove any remaining \yb{174} atoms with a resonant ``push'' pulse on the broad $\sS{0} \rightarrow \sP{1}$ transition.
After evaporative cooling, typically $3\text{-}5 \times 10^3$ \yb{171} atoms in each of the two nuclear spin states $m_F \in \{-1/2, +1/2\}$ remain in the optical dipole trap at a temperature of $T \simeq 0.15T_F$, which is determined by fitting density profiles to in-situ absorption images.
Here, $T_F$ is the Fermi temperature.
We note that the extremely small scattering length $-2.8(3.6)a_0$ of \yb{171} in the ground state~\cite{kitagawa:2008} generally leads to a particularly long thermalization time scale.
Here, $a_0$ is the Bohr radius.

\begin{figure}[t]
  \includegraphics[width=\columnwidth]{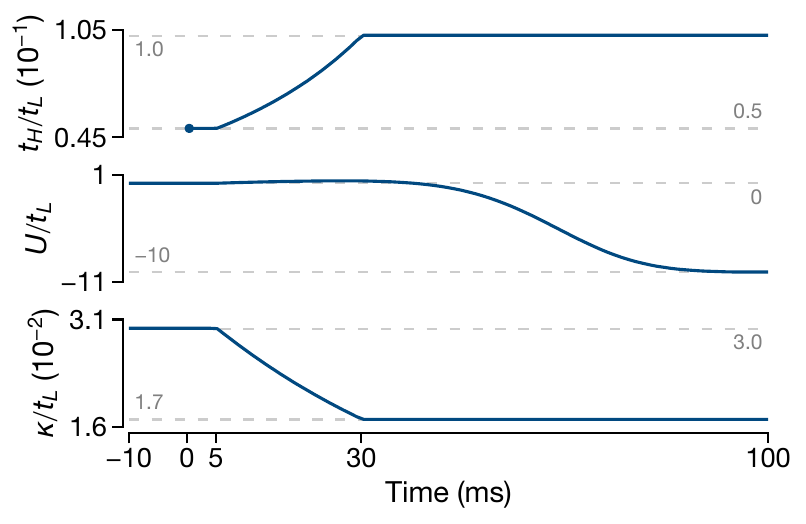}
  \caption{\label{fig:exp-seq}%
    Time dependence of the Hubbard parameters $t_H$, $U$, and $\kappa$ during the state preparation for the target values $t_H/t_L \simeq 0.1$ and $U/t_L \simeq -10$.
    All parameters are estimated from the voltages controlling the lattice depth and magnetic field strength in the experiment.
    Time is given relative to the clock-laser excitation pulse which quenches the hopping ratio from $t_H/t_L = 1$ (not visible on the chosen scale) to $t_H/t_L \simeq 0.05$, indicated with a small circle.
  }
\end{figure}

The atoms are loaded into the optical lattices sequentially with s-shaped ramps, and we first ramp up the vertical state-independent lattice to \mbox{$\approx 30\,E_\text{rec}^m$} within $120\ms$.
Then, the state-independent lattice along the other axis is ramped up to the same depth within~$300\ms$.
After subsequently loading the atoms into the \mbox{$\approx 7\Erec$} deep state-dependent lattice (SDL) within~$300\ms$, we proceed with the state preparation (also see main text), which includes changing the SDL depth within~$25\ms$ (linear ramp) and the magnetic field within~$70\ms$ (s-shaped ramp) to set the desired Hubbard parameters.
We show the time dependence of these values for an exemplary case in Fig.~\ref{fig:exp-seq} and also refer to Table~\ref{tbl:exp-params}, which summarizes the experimental parameters for all figures in the main text.
%
\setlength{\tabcolsep}{0pt}
\begin{table*}
  \caption{\label{tbl:exp-params}%
    Experimental parameters for each figure presented in the main text.
    The lattice depth is given for the light species [$V_H = 3.06(4) V_L$ for the heavy species unless noted otherwise, see Section~\ref{sec:sdl}].
    We also show next-nearest neighbor and perpendicular hopping amplitudes to illustrate the accuracy of a one-dimensional tight-binding description.}\vspace{0.5em}
  \begin{ruledtabular}
  \begin{tabular}{l@{\hskip 1em} r@{ -}l@{\hskip 0.5em} r@{ -}l@{\hskip 0.5em} r@{ -}l@{\hskip 0.5em} r@{ -}l@{\hskip 0.5em}}
    & \multicolumn{2}{c}{Fig.~2(a)} & \multicolumn{2}{c}{Fig.~3} & \multicolumn{2}{c}{Fig.~4(a)} & \multicolumn{2}{c}{Fig.~4(b)\footnote{For this measurement, we operate the state-dependent lattice at the wavelength $\lambda^\prime \simeq 690.1\nm$ such that $V_H= 1.97(5)V_L$.}}\\\hline
    Total number of light atoms $N_\text{tot}$ $\left(10^3\right)$ & \multicolumn{2}{c}{$4.2(3)$} & \multicolumn{2}{c}{$4.5(3)$} & \multicolumn{2}{c}{$3.4(3)$} & \multicolumn{2}{c}{$3.8(6)$} \\
    Effective number of light atoms per tube $\mathcal{N}[\Delta \mathcal{N}]$ & \multicolumn{2}{c}{$19[8]$} & \multicolumn{2}{c}{$19[9]$} & \multicolumn{2}{c}{$17[7]$} & \multicolumn{2}{c}{$18[8]$} \\[0.5em]
    Lattice depth $V_L$ $\left(\Erec\right)$ & $0.69(3)$ & $11.8(5)$ & \multicolumn{2}{c}{$4.7(2)$} & \multicolumn{2}{c}{$4.7(2)$} & \multicolumn{2}{c}{$4.2(3)$} \\
    Hopping amplitude $t_L$ $\left(h \times \text{Hz}\right)$ & $33(4)$ & $488(2)$ & \multicolumn{2}{c}{$184(10)$} & \multicolumn{2}{c}{$182(8)$} & \multicolumn{2}{c}{$187(12)$} \\[0.5em]
    Next-nearest neighbor hopping $t_L^{(2d)}/t_L$ $\left(10^{-2}\right)$ & $0.75(10)$ & $21.8(2)$ & \multicolumn{2}{c}{$5.7(4)$} & \multicolumn{2}{c}{$5.6(3)$} & \multicolumn{2}{c}{$6.2(5)$} \\
    Perpendicular hopping $t_\perp/t_L$ $\left(10^{-3}\right)$ & $2.1(3)$ & $31(8)$ & \multicolumn{2}{c}{$6(1)$} & \multicolumn{2}{c}{$6(1)$} & \multicolumn{2}{c}{$6(1)$} \\[0.5em]
    Hopping ratio $t_H/t_L$ & \multicolumn{2}{c}{n/a} & \multicolumn{2}{c}{$0.104(7)$} & \multicolumn{2}{c}{$0.102(6)$} & \multicolumn{2}{c}{$0.34(3)$} \\
    Interaction strength $U/t_L$ & \multicolumn{2}{c}{n/a} & $-19(2)$ & $4.6(3)$ & $-9.7(7)$ & $0.02(4)$ & $-10(1)$ & $0.01(5)$ \\
    Harmonic confinement $\kappa/t_L$ $\left(10^{-2}\right)$ & $0.63(3)$ & $9(1)$ & \multicolumn{2}{c}{$1.7(1)$} & \multicolumn{2}{c}{$1.7(1)$} & \multicolumn{2}{c}{$1.7(2)$} \\
    \end{tabular}
  \end{ruledtabular}
\end{table*}

We use a linearly polarized clock laser beam to excite atoms from the $\ket{\sS{0},m_F = +1/2}$  ground state (denoted $\ket{g\uparrow}$) to the $\ket{\tP{0},m_F = +1/2}$ clock state (denoted $\ket{e\uparrow} \equiv \ket{H}$).
This laser beam co-propagates with one of the deep perpendicular lattices to drive the transition in the Lamb-Dicke regime.
To overcome the inhomogeneous broadening present in the SDL and to quench the hopping amplitude for the heavy species, we use a high-intensity excitation pulse with Rabi frequency $\Omega \simeq 2\pi \times 3\kHz$.
We address singly and doubly occupied lattice sites with this single-frequency pulse by applying a large magnetic field $\simeq 1533\Gauss$ corresponding to the zero crossing of the orbital Feshbach resonance (see Section~\ref{sec:ofr}), where the interaction shift of doubly occupied lattice sites vanishes.
In general, the fidelity of the excitation pulse is fairly good, with $> 95\%$ of the atoms driven to $\ket{H}$.
We note that the residual number of remaining $\ket{g\uparrow}$ atoms only interact with atoms in $\ket{H}$ (interaction strength $U_{H,g\uparrow} \gg t_L$).
However, this is independent of the chosen heavy-light interaction strength and should only have a minor influence on our observations.

After quenching the hopping ratio $t_H/t_L$ of the non-interacting mixture, we slowly ramp the interaction strength towards the desired value by varying the magnetic field accordingly (see Fig.~\ref{fig:exp-seq}).
In general, this process can significantly change the initial state, particularly the fraction of doublons as the Hamiltonian is gradually modified during the ramp of the interaction.
However, numerical simulations suggest that this fraction only weakly depends on the chosen $U$ parameter [see Section~\ref{sec:initial-state} for details].
We also verify this experimentally by determining an estimate of the doublon fraction with clock-line spectroscopy.
After the state-preparation protocol (see Fig.~\ref{fig:exp-seq}), we freeze the motion of both species by rapidly ramping up the SDL to $V_L\approx 15\Erec$.
Subsequently, we ramp the magnetic field to $B\simeq1450\Gauss$ and employ clock-line spectroscopy to determine the amplitude ratio of the signal originating from singly and doubly occupied lattice sites.
To circumvent the inhomogeneous broadening in the SDL and increase the signal, we only consider a central region of the atomic cloud containing~$\approx 5\%$ of atoms.
In this way, we determine a maximum change of the doublon fraction~$\approx16\%$ between~$U/t_L\approx -20$ and~$4$, which is qualitatively consistent with the numerical simulations shown in Fig.~\ref{fig:sim-state-prep}.

\subsection{Lattice loading}\label{sec:lat-loading}
For the theoretical description of the experimental procedure (see Section~\ref{sec:theory}), the distribution of atoms across the different one-dimensional systems (tubes) is of central importance.
Since our absorption imaging inherently averages across one axis, we cannot determine this quantity directly.
Instead, we estimate it from parameters in the dipole trap, together with the assumption of constant entropy during the process of lattice loading~\cite{koehl:2006}.
This assumption only approximately holds as we can detect an increase of the temperature (and entropy) by \mbox{$\approx 30\%$} when loading the atoms from the lattice back into the dipole trap.
We also simplify the lattice loading process by assuming the state-independent lattices are ramped up simultaneously (see Section~\ref{sec:state-prep}).

\begin{figure}[t]
  \includegraphics[width=\columnwidth]{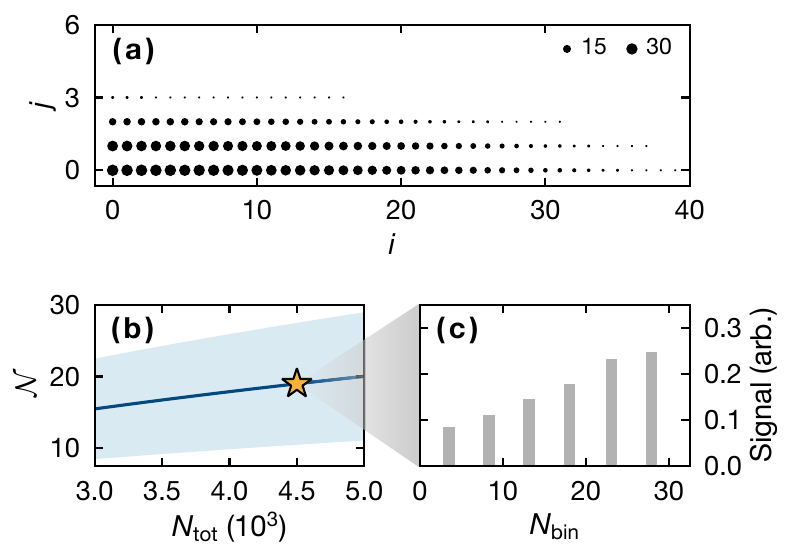}
  \caption{\label{fig:tube-distr}%
    Distribution of atoms across different one-dimensional systems (tubes).
    (a)~Typical distribution of $N_\text{tot}=4.5 \times 10^3$ (light) atoms across the tubes generated by the perpendicular state-independent lattices.
    Each black circle represents a tube with the size indicating the number of atoms (see legend).
    We only plot a single quadrant due to the symmetry of the distribution.
    (b)~Effective number of light atoms per tube $\mathcal{N}$ for variable total light atom numbers $N_\text{tot}$.
    The blue-shaded area corresponds to the region of $\mathcal{N} \pm \Delta \mathcal{N}$.
    (c)~Approximate distribution of light atoms across different bins for $N_\text{tot}=4.5 \times 10^3$ (yellow star) and their contribution to the measured signal.
    This distribution is obtained by averaging over multiple relative phases $-0.5 \leq \varphi_i, \varphi_j < 0.5$ [see Eq.~\eqref{eq:tube-chem-pot}].
    }
\end{figure}

From the total number of light atoms $N_\text{tot}$, the trapping frequencies $(\omega_x, \omega_y, \omega_z) = 2\pi\times[38(1), 35(1), 402(1)] \Hz$ and temperature $T \simeq 0.15 T_F$ in the optical dipole trap, we determine the total entropy $S(N_\text{tot}, T) \simeq N_\text{tot} \times 1.4 k_B$ of the initial state.
For the final state in the array of tubes formed by the perpendicular lattices, we fit the chemical potential $\mu_0$ and temperature $T^\prime$ as parameters of the total entropy,
\begin{align}
  S^\prime(\mu_0, T^\prime) = \sum_{i,j} S_{ij}(\mu_0 + \mu_{ij}, T^\prime),
\end{align}
under the constraint of constant entropy, $S = S^\prime$, and conservation of atom number, $\sum_{i,j} N_{ij} = N_\text{tot}$.
Here, $S_{ij}$ is the entropy and $N_{ij}$ the number of atoms in the $(i, j)$-th tube, described by   a harmonically trapped one-dimensional Fermi gas.
The chemical potential $\mu_{ij}$ accounts for the local offset of the harmonic potential from the optical dipole trap,
\begin{align}\label{eq:tube-chem-pot}
  \mu_{ij} = \frac{1}{2} m d_m^2 \left[\omega_y^2 {(i + \varphi_i)}^2 + \omega_z^2 {(j + \varphi_j)}^2\right]
\end{align}
with $m$ the mass of a \yb{171} atom, $d_m = \lambda_m / 2$ the spacing of the state-independent lattice, and $\varphi_i$, $\varphi_j$ corresponding to the relative phase of the optical lattice and optical dipole trap.
In Fig.~\subfigref{tube-distr}{a}, we plot the resulting distribution of $\{N_{ij}\}$ for typical parameters in our experiment.
From this distribution, we also calculate the effective quantities \mbox{shown in Table~\ref{tbl:exp-params} and Fig.~\subfigref{tube-distr}{b}},
\begin{align}\label{eq:tube-atom-number}
  \mathcal{N} &= \frac{1}{N_\text{tot}} \sum_{i,j} N_{ij}^2\text{, and} \\
  \Delta\mathcal{N} &= {\left[ \frac{1}{N_\text{tot}} \sum_{i,j} N_{ij} {(\mathcal{N} - N_{ij})}^2 \right]}^{1/2}.
\end{align}
For our theoretical calculations (see Section~\ref{sec:non-int-calc}), we distribute $\{N_{ij}\}$ over six bins such that the effective width of each bin is approximately constant and average over ten relative phases for each $\varphi_i$ and $\varphi_j$ [see Fig.~\subfigref{tube-distr}{c}].
We only consider the \mbox{$\approx 300$} significantly filled tubes ($N_{ij} \geq 1$) in this procedure, which account for more than $98\%$ of the signal.

Finally, we calculate all relevant parameters in the state-dependent lattice using the atom numbers $\{N_{ij}\}$ and entropies $\{S_{ij}\}$, which we increase by $30\% / 2 = 15\%$ to account for half of the temperature increase observed when loading back into the optical dipole trap.
From this estimate and the exact diagonalization of the single-particle Hamiltonian in Eq.~\eqref{eq:single-particle-ham}, we find an effective filling $n \simeq 0.5$, fraction of doublons $\mathcal{D} \simeq 0.3$, and temperature $T \simeq 2 t_L/k_B$ for a typical total number of light atoms $N_\text{tot} = 4.5\times10^3$ and lattice depth $V_L=4.7\Erec$.
In addition, we estimate the effective system size $l \approx 30$ lattice sites of a typical tube from the root-mean-square width of the density distribution along the one-dimensional system, which is also obtained with exact diagonalization techniques~(see Section~\ref{sec:non-int-calc}).

To verify our modeling of the lattice loading process, we compare the in-situ column density to a theoretical prediction obtained from the atom numbers and entropies in each tube.
Here, we find qualitative agreement with deviations most likely originating from the finite imaging resolution and imperfect confinement potentials.

\subsection{Absorption imaging}
We perform in-situ absorption imaging of the atomic cloud on the broad $\sSF{0}{1/2} \rightarrow \sPF{1}{3/2}$ transition, which allows us to directly determine the column density of the light atoms.
In the experiment, we freeze the atomic density by rapidly increasing the lattice depth to \mbox{$\approx15\Erec$} within $2.5\ms$.
Then, we lower the magnetic field within $30\ms$ to $2\Gauss$ and image the light atoms with a $5\us$ long high-intensity pulse.
To calibrate the atomic cross-section and the absolute atom number, we use a technique analog to the one used for \yb{173} in Ref.~\cite{darkwahoppong:2019}.

Finally, we briefly comment on the indirect imaging of the heavy atoms, which are detected after the first imaging pulse has removed all light atoms in the \sS{0} state.
We employ a $3\ms$ long laser pulse resonant with the $\tPF{0}{1/2} \rightarrow \tDF{1}{3/2}$ transition to repump all clock state (heavy) atoms back to the ground state.
Subsequently, these atoms are detected with absorption imaging on the same transition as the light atoms.
However, this detection technique does not reliably determine the in-situ density of the heavy atoms for several technical reasons.
First, the repumping introduces motional blurring and additional atom loss up to $30\%$~\cite{bettermann:2020}, which depends significantly on the local density.
Second, on-site heavy-light atom pairs can either be projected onto the repulsive branch or are continuously transferred into the deeply bound molecular state~\cite{bettermann:2020} during the relatively quick ramp of the magnetic field before the imaging.
The exact process strongly depends on the technical details of the ramp, and our data suggests that heavy atoms transferred to the molecular state are dark for our detection method, thereby introducing additional loss on the order of \mbox{$\sim 10\%$}.
Although the generally \mbox{$\approx 40\%$} lower number of detected heavy atoms [see Figs.~\subfigref{density-dev-loss}{c} and~\subfigref{density-dev-loss}{d}] is approximately consistent with our estimates, the measured column densities still contain systematic errors, and we therefore do not use them for any quantitative analysis.

\subsection{Hubbard parameters}
We determine the hopping amplitudes $t_L$ and $t_H$ from two separate tight-binding band structures for light and heavy atoms~\cite{bloch:2008}.
The on-site interaction $U$ is determined from the overlap of the Wannier functions $w_\alpha(x)$ and the magnetic field-dependent $s$-wave scattering length $a$ (see Section~\ref{sec:ofr}),
\begin{align}\label{eq:hubbard-int}
  U = \frac{4\pi\hbar^2 a}{m}  \int d\mathbf{r}\, {\left|w_L(x)\right|}^2 \, {\left| w_H(x)\right|}^2 \,{\left| w_\perp(y, z) \right|}^4.
\end{align}
Here, $\mathbf{r} = (x, y, z)$, $m$ is the mass of a \yb{171} atom, and $w_\perp(y, z) = w_y(y) w_z(z)$ is the combined Wannier function along the perpendicular axes $y$ and $z$.
The main experimental input parameters for the band-structure calculations are the lattice wavelength and depth.
The former is fixed with high precision by locking the lasers to a wavelength meter using a digital servo.
In contrast, the latter is calibrated for each measurement using deep lattices to minimize systematic uncertainties.
For this calibration, we employ lattice modulation spectroscopy~\cite{friebel:1998} for one of the state-independent lattice axes and resolved sideband spectroscopy on the clock transition for the remaining two lattice axes (state-independent and state-dependent, also see Section~\ref{sec:sdl}).
From these measurements, we also estimate the uncertainties of the Hubbard parameters.
Finally, the value of the confinement parameter $\kappa = m\omega^2 d^2$ is obtained by measuring the trapping frequency $\omega$ after suddenly displacing the trap center and fitting the frequency of the resulting center-of-mass oscillations.
We perform this measurement with the state-dependent lattice turned off as it does not significantly contribute to the confinement.

\subsection{Orbital Feshbach resonance}\label{sec:ofr}
\begin{figure}[t]
  \includegraphics[width=\columnwidth]{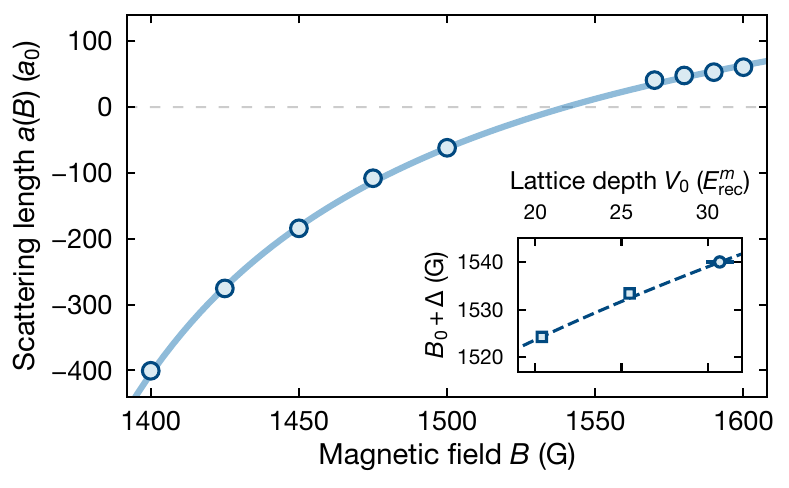}
  \caption{\label{fig:ofr}%
    Scattering length extracted from clock-line spectroscopy of the single- and two-particle resonance in a nearly isotropic three-dimensional state-independent optical lattice with depth $V_0=30.7(8) E^m_\text{rec}$.
    The circles correspond to individual data points and the solid line is a fit to Eq.~\eqref{eq:scat-len}.
    We do not plot data points close to $a(B)=0$ since the resonances overlap in this region making the fit of the experimental data prone to systematic errors.
    In the inset, we show the zero crossing ($B_0 + \Delta$) in an isotropic state-independent optical lattice with variable depth.
    Here, the circle corresponds to the data set in the main panel whereas the squares represent separate measurements.
    The dashed line shows the expected scaling from Eq.~\eqref{eq:zero-crossing} relative to the data point $V_0=30.7 E^m_\text{rec}$.
  }
\end{figure}

Details on the orbital Feshbach resonance (OFR) in \yb{171} can be found in Ref.~\cite{bettermann:2020}.
Here, we only briefly discuss how we employ clock-line spectroscopy on the single-particle and two-particle resonance to measure the magnetic field-dependent $s$-wave scattering length between $\ket{e\uparrow}\equiv \ket{H}$ and $\ket{g\downarrow} \equiv \ket{L}$ atoms.
This measurement is performed in a nearly isotropic three-dimensional state-independent optical lattice with depth \mbox{$\approx 30.7(8)\Erec$} to avoid the inhomogeneous broadening present in the state-dependent lattice (SDL).
From the transition energies at various magnetic fields $B$, we extract the corresponding scattering length $a(B$) with an expression similar to Eq.~\eqref{eq:hubbard-int}.
In Fig.~\ref{fig:ofr}, we show the experimental data together with a fit to the following general form~\cite{chin:2010},
\begin{align}\label{eq:scat-len}
  a(B) = a_\text{bg} \left[1 - \frac{\Delta}{(B - B_0)} \right].
\end{align}
Here, the fit parameters are $a_\text{bg} = 333(14) a_0$, $\Delta = 255(6)\Gauss$, and $B_0 = 1285(5)\Gauss$ with $a_0$ the Bohr radius.

We find a significant dependence of the zero crossing $B_{a=0} = B_0 + \Delta$ (calculated from the fitted parameters) on the chosen lattice depth (see inset of Fig.~\ref{fig:ofr}).
The effective shift of $B_{a=0}$ most likely originates from the energy dependence of the open-channel scattering amplitude~\cite{laird:2020}.
For the lattice depth $V_0$, the zero-point energy of the ground band is approximately given by
\begin{align}\label{eq:zero-crossing}
  \epsilon_0 \simeq 3 E_\text{rec}^{m}\left[ {\left(\frac{V_0}{E_\text{rec}^{m}}\right)}^{1/2}  - \frac{1}{4}\right]
\end{align}
and the resulting shift of the zero crossing can be obtained from $\delta B_{a=0} = \epsilon_0 / |\delta u|$ with $\delta u = - h \times 399.0(1) \Hz$ the differential Zeeman shift of the open and closed channel~\cite{laird:2020,bettermann:2020}.
In this simple approximation, we neglect any energy dependence of the scattering length itself but still find good agreement between experiment and theory.
The situation for the SDL is slightly different, and we have to consider the mean zero-point energy of a light and a heavy atom in combination with different optical lattices.
We utilize the symmetry (with respect to $U=0$) of the signal in Fig.~3 and benchmark our approach with additional measurements for different depths of the SDL [see Fig.~\subfigref{transport}{c}].
Since we do not know the precise location of the zero crossing without the trap, we extract it from a fit to this data.

Finally, we briefly comment on the lifetime of the heavy-light mixture close to the OFR, which we can monitor up to a maximum duration of $400\ms$ in our experiment.
In general, losses over this time scale are relatively small for the regime $1450\text{-}1600\Gauss$ but increase quickly for magnetic fields \mbox{$\lesssim 1450\Gauss$}.
Importantly, these losses should be negligible for most magnetic fields employed for the measurements in the main text [also see Figs.~\subfigref{density-dev-loss}{c} and~\subfigref{density-dev-loss}{d}].

\vspace{2\baselineskip}
\subsection{State-dependent lattice}\label{sec:sdl}
Our state-dependent optical lattice (SDL) is generated by retro-reflecting a monochromatic laser beam with wavelength $\lambda = 671.509(1)\nm$, almost identical to the realization in Ref.~\cite{riegger:2018}.
At this particular wavelength, the polarizability $\alpha$ of $\ket{g}$ (light) and $\ket{e}$ (heavy) atoms differs significantly, giving rise to different lattice depths and hopping amplitudes. 
We determine the corresponding polarizability ratio $p=\alpha_e / \alpha_g = V_H / V_L$ with interleaved sideband spectroscopy on the clock transition.
Here, $V_H$ and $V_L$ are the lattice depths experienced by heavy ($\ket{e}$) and light ($\ket{g}$) atoms, respectively.

For this measurement, we use a separate clock laser beam co-aligned with the SDL axis to change the orbital state from $\ket{g}$ to $\ket{e}$ (or vice versa) and simultaneously the band index from $m$ to $m^\prime$.
\begin{figure}[t]
  \includegraphics[width=\columnwidth]{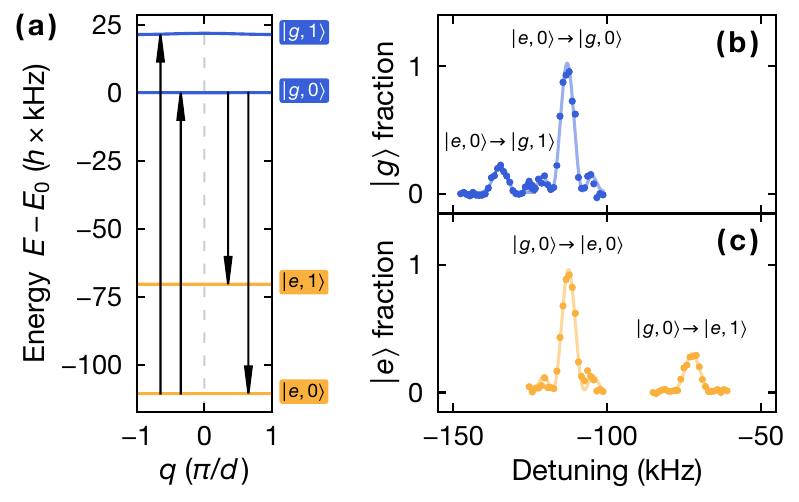}
  \caption{\label{fig:pol-ratio}%
    Measurement of the polarizability ratio $p$ in the state-dependent lattice (SDL).
    (a)~Band structure of a $22.6(6)\Erec$ deep SDL with the two lowest-lying energy bands of $\ket{g}$ atoms (blue) and $\ket{e}$ atoms (yellow).
    The four transitions required to determine the $\ket{g}$ and $\ket{e}$ band gap are shown as black arrows at exemplary quasi-momenta $q$.
    All energies are shown relative to $E_0$, which is the mean energy of the $\ket{g}$ ground band.
    \mbox{(b),(c)}~Clock-line spectra of the carrier and sideband transitions starting from (b)~$\ket{e,0}$ and (c)~$\ket{g, 0}$ for variable detuning of the clock laser.
    Here, the circles correspond to data points and the solid lines to individual fits of the line shapes.
  }
\end{figure}
We perform two sets of measurements, one with the atoms initially prepared in $\ket{e, m=0}$ and the other with the atoms in $\ket{g, m=0}$.
By driving the carrier ($m=0 \rightarrow m^\prime=0$) as well as the first sideband transition ($m=0 \rightarrow m^\prime=1$), we find two transition frequencies by fitting the line shapes as shown in Figs.~\subfigref{pol-ratio}{b} and~\subfigref{pol-ratio}{c}.
Subtracting the two frequencies yields the band gap individually for $\ket{g}$ and $\ket{e}$ atoms, which allows us to determine the corresponding lattice depths $V_L$ and $V_H$ from a band-structure calculation.
With these results, we find a polarizability ratio $p=3.06(4)$ from multiple measurements.
By employing the same technique, we also determine the polarizability ratio $p^\prime = 1.97(5)$ for the SDL wavelength $\lambda^\prime = 690.076(1)\nm$.

The much larger polarizability $\alpha_e$ of the clock state $\tP{0}$ ($\ket{e}$) arises since the SDL is operated only \mbox{$\approx 20\nm$} detuned from the $\tP{0}\rightarrow\tS{1}$ transition wavelength.
Consequently, the scattering of lattice photons becomes significant and heavy atoms off-resonantly excited to the \tS{1} state quickly decay back to either of the lower-lying \tP{J=0,1,2} states with a ratio of $15{:}40{:}45$ according to the Clebsch-Gordan coefficients.
Thus, the scattering of only two photons is sufficient to almost completely deplete the $\ket{e}$ state, and the dominant dissipation in the system is the loss of heavy atoms.
More precisely, heavy atoms are converted with almost equal probability to either \tP{1} atoms with a subsequent quick decay back to $\ket{g\uparrow}$ or \tP{2} atoms, which are most likely quickly lost from the trap.

We characterize the loss dynamics by monitoring the remaining number of heavy atoms after preparing a pure sample in $\ket{e\uparrow} = \ket{H}$.
For the relevant lattice depths ($3\text{-}7\Erec$), we find $1/e$ lifetimes~\mbox{$\sim 10^3 \hbar/t_L$} obtained by fitting the experimental data.
While these loss rates are negligible for probing the transport directly after the displacement of the trap minimum in Fig.~3, the total loss becomes appreciable with \mbox{$\approx 20\%$} for the longest observation times \mbox{$\simeq 400\hbar/t_L$} of the density dynamics in~Fig.~4.

\begin{figure*}[t]
  \includegraphics[width=\textwidth]{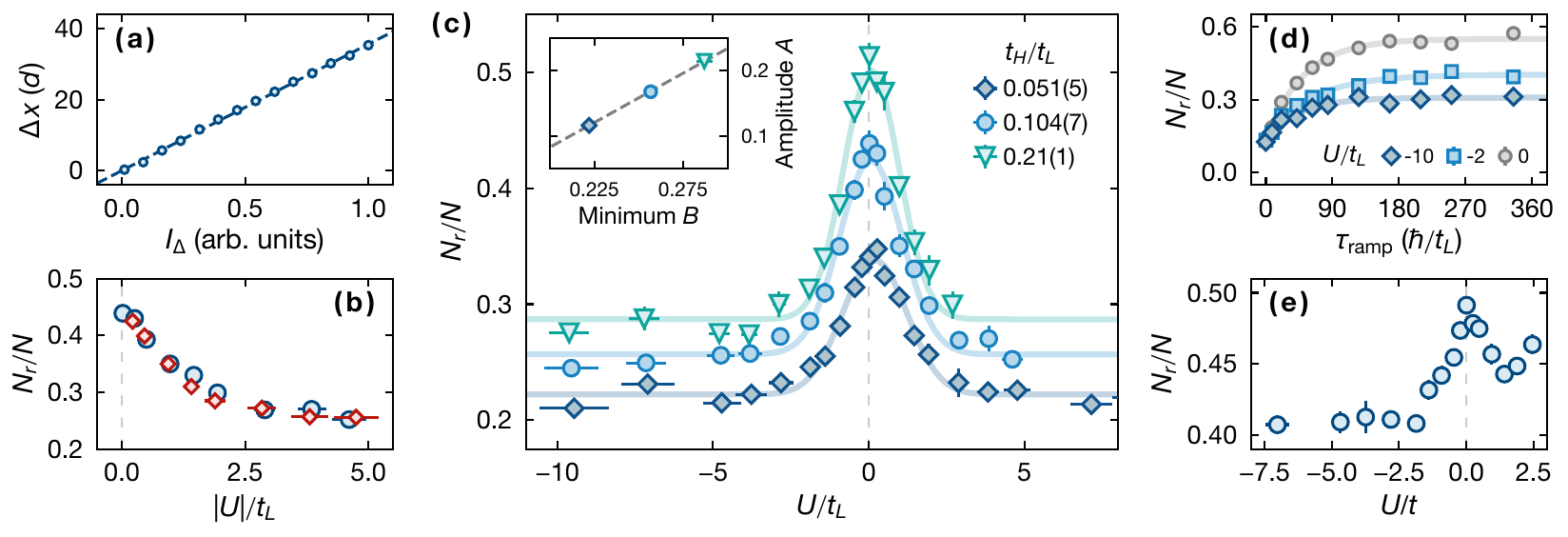}
  \caption{\label{fig:transport}%
    Details on the displacement of the trap center and the transport measurement.
    (a)~Transport distance $\Delta x$ extracted from a Gaussian fit of the atomic column density for variable intensity $I_\Delta$ in the off-centered dipole trap.
    The dashed line corresponds to a linear fit.
    (b)~Transported fraction $N_r/N$ from Fig.~3 shown for $|U|/t_L$ to illustrate the symmetry between attractive ($U < 0$, red diamonds) and repulsive ($U > 0$, blue circles) interactions.
    (c)~Transported fraction $N_r/N$ for variable hopping ratio $t_H/t_L$ and interaction strength $U/t_L$.
    Note that the $t_H/t_L = 0.104(7)$ data set (light blue circles) corresponds to the data shown in Fig.~3.
    The solid lines are fits to $A \exp[-{(U/t_L)}^2/(2\sigma^2)] + B$ for extracting the amplitude~$A$ and minimum~$B$ shown in the inset together with a linear fit (dashed gray line).
    (d)~Transported fraction $N_r/N$ for $t_H/t_L=0.102(6)$, $\Delta x \simeq 43(3)d$, and variable ramp duration $\tau_\text{ramp}$.
    The solid lines are a guide to the eye.
    (e)~Transported fraction $N_r/N$ for the special case $t \equiv t_H = t_L$ and variable interaction strength $U/t$ with the transport distance $\Delta x \simeq 44(3)d_m$ and ramp duration $94(3)\hbar/t$.
  }
\end{figure*}

\subsection{Displacement of the trap center}\label{sec:displ-trap}
We use an off-centered state-independent dipole trap beam operated at the magic wavelength to displace the minimum of the trap.
The spatially-dependent potential resulting from this approximately Gaussian laser beam is given as
\begin{align}\label{eq:dip-trap-pot}
  U_\Delta(x) = \frac{1}{2} m \omega_\Delta^2 {(x-x_\Delta)}^2 \left\{ 1 + \mathcal{O}\left[ \frac{{(x-x_\Delta)}^2}{w_0^2} \right] \right\}.
\end{align}
Here, $m$ is the  mass of a \yb{171} atom, $\omega_\Delta \in 2\pi \times [0, 23] \Hz$ is the trapping frequency proportional to the chosen intensity of the laser beam, $x_\Delta \approx 50\um$ is the displacement of the beam with respect to the initial trap minimum $i_0$ (see main text), and $w_0 \approx 1.2 x_\Delta$ is the estimated waist of the Gaussian laser beam. 
Neglecting the higher-order corrections to $U_\Delta(x)$, the shifted location of the trap center $x_0$ is given as
\begin{align}
  \Delta x = x_\Delta \frac{\omega_\Delta^2}{\omega^2 + \omega_\Delta^2}
\end{align}
with $\omega = 40(1)\Hz$ (see main text).
For our measurements discussed in the main text, the intensity $I_\Delta \propto \omega_\Delta^2$ of the off-centered dipole trap is linearly increased, such that the trap minimum is displaced accordingly.
We note that the terms in Eq.~\eqref{eq:dip-trap-pot} also modify the effective trapping frequency $\omega$, which we estimate to be on the order of \mbox{$\sim 10\%$} and neglect in our theoretical description.

We calibrate the final value of $\Delta x$ independently by ramping up the dipole trap before loading the state-dependent lattice.
This allows us to extract the trap center and effective displacement from a Gaussian fit of the atomic column density.
In Fig.~\subfigref{transport}{a}, we show the result of such fits for variable intensity of the optical dipole trap.
We estimate the statistical uncertainty of $\Delta x$ conservatively with \mbox{$\approx 3d$} from the observed $2\sigma$-fluctuations of the trap center over typical measurement durations.

\begin{figure*}[t]
  \includegraphics[width=\textwidth]{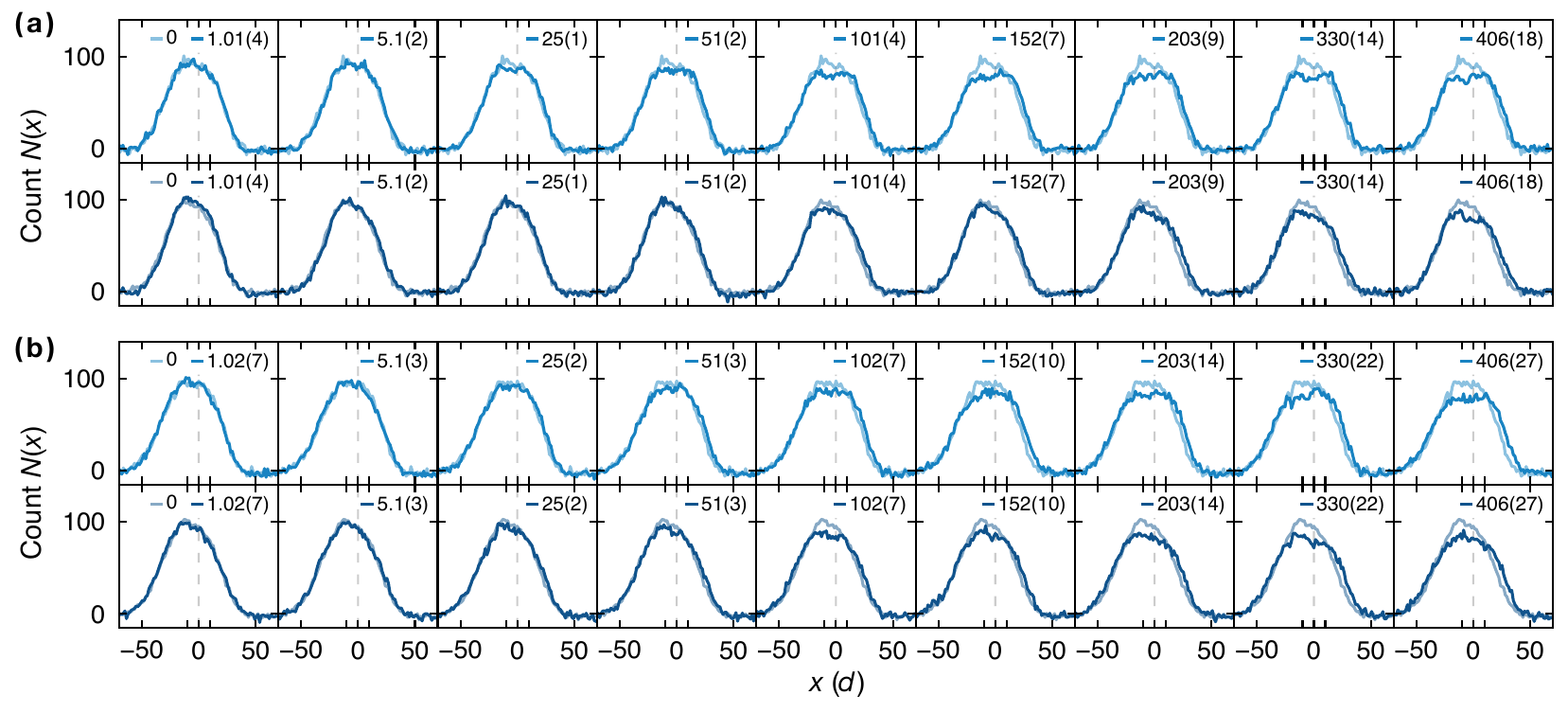}
  \caption{\label{fig:int-densities}%
     One-dimensional densities for the hopping ratios (a)~$t_H/t_L=0.102(6)$ as well as (b)~$0.34(3)$ and multiple hold times as indicated in the top right legends (units of $\hbar/t_L$).
     Each row corresponds to either one of the interaction strengths $U/t_L = -1.9(3)$ (light blue, top row of each panel) and~$-10(1)$ (dark blue, bottom row of each panel).
  }
\end{figure*}
\subsection{Transport measurement}
In this section, we discuss details and show additional experimental data for the transport measurements.
As discussed in the main text, we observe a mostly symmetric signal (see Fig.~3), which only depends on the absolute value of the interaction strength $|U|/t_L$, shown explicitly in Fig.~\subfigref{transport}{b}.
Formally, the dynamical symmetry of our model in the limit $\kappa=0$~\cite{schneider:2012} is broken by the finite confinement.
Due to the relatively small energy scale of $\kappa$, we nevertheless still expect the dynamics to be comparable for attractive and repulsive interactions, which is confirmed by our numerical simulations shown in Fig.~3 [also see Section~\ref{sec:transport-sim}].

In Fig.~\subfigref{transport}{c}, we show additional measurements for the hopping ratios $t_H/t_L \simeq 0.05$ and $0.2$ obtained by adjusting the depth of the state-dependent lattice (SDL) accordingly.
We note that this also changes the relative confinement to $\kappa/t_L=2.9(3)\times 10^{-2}$ and $1.1(1)\times 10^{-2}$ compared to the case of $t_H/t_L\simeq 0.1$ discussed in the main text.
In general, we observe a similar functional dependence on the interaction parameter, but the overall amplitude and value for $U/t_L \simeq 0$ is reduced for smaller hopping ratios, which we attribute to a larger fraction of atoms occupying Wannier-Stark-localized states.
The inset in Fig.~\subfigref{transport}{c} illustrates the approximate linear behavior, which we find between the amplitude and minimum of a Gaussian fit.

To determine the optimal speed for the translation of the trap minimum in our measurement, we also characterize the transported fraction $N_r/N$ for constant ramp distance $\Delta x$ and variable ramp duration $\tau_\text{ramp}$. 
For the probed interaction parameters $U/t_L \simeq 0$, $-2$, $-10$, we find a monotonous increase of the observable, which approximately follows an exponential saturation curve, as shown in Fig.~\subfigref{transport}{d}.
For the measurements in the main text, we chose the ramp speed \mbox{$\approx 0.5d t_L/\hbar$}, which corresponds to approximately twice the $1/e$ time in the non-interacting case.
We note that the transported fraction $N_r/N$ remains significantly reduced for all finite interactions and independent of the exact ramp duration.

To complement the data in Fig.~3, we perform an additional measurement for a mass-balanced configuration with $t \equiv t_L = t_H$.
This is achieved by replacing the SDL with a state-independent lattice operated at the magic wavelength with the same lattice depth for the $\ket{L}$ as well as $\ket{H}$ state.
We emphasize that this measurement only serves as a qualitative comparison since the beam parameters of the two lattices differ strongly, which introduces systematic differences.
In general, we observe a much smaller suppression of the transported fraction and a large asymmetry between attractive ($U < 0$) and repulsive ($U > 0$) interactions [see Fig.~\subfigref{transport}{e}].
In contrast to the situation in the SDL, the fraction of doublons here strongly depends on the interaction parameter due to the nearly adiabatic state preparation.
Thus, the inhibited transport originates most likely from the slow dynamics \mbox{$\propto t^2/U$} of doublons in the system, as similarly observed in Ref.~\cite{strohmaier:2007}.
We verify the increased (reduced) doublon fraction for attractive (repulsive) interactions with clock-line spectroscopy and find a modulation of up to $\approx 50\%$---in reasonable agreement with the observed asymmetric lineshape.

\subsection{Density dynamics}\label{sec:dens-dev}
We characterize the non-equilibrium dynamics after the displacement of the trap minimum for variable hold times by recording the density deviation,
\begin{align}\label{eq:density-dev}
  \delta n{(\tau)} = {\left\{ {\int dx\,n(x, \tau) {[ n(x, \tau) - n(x, 0) ]}^2} \right\}}^{(1/2)},
\end{align}
as discussed in the main text.
The effective weight $n(x, \tau)$ helps to significantly reduce the contribution of noise in the integral.
The one-dimensional density~$n(x, \tau)$ is calculated from the two-dimensional atomic column density $\mathrm{CD}(x, y; \tau)$ by normalization and integration along $y$,
\begin{align}
  n(x, \tau) = \frac{1}{\int dx dy \, \mathrm{CD}(x, y; \tau)} \int dy\, \mathrm{CD}(x, y; \tau).
\end{align}
The relaxation of $n (x, \tau)$ shown explicitly in the two right panels of Fig.~4(a) for $t_H/t_L \simeq 0.1$ and $U/t_L \simeq -10$,~$0$ can also be observed for other interaction parameters and hopping ratios, which are shown in Fig.~\ref{fig:int-densities}.
Compared to smaller interaction strength or larger hopping ratio, the one-dimensional densities for $t_H/t_L\simeq 0.1$ and \mbox{$U/t_L \simeq -10$} appear to not show significant relaxation for short hold times $\tau \lesssim 50 \hbar/t_L$.

Importantly, the signatures of $\delta n (\tau)$ discussed in the main text can also be similarly found in the fraction of transported atoms $N_r/N$.
However, this observable is particularly sensitive to the exact calibration of the trap minimum $i_0$ (see Section~\ref{sec:displ-trap}) as the atomic cloud does not fully separate for the relatively short transport distance $\Delta x \simeq 20d$ in this measurement.
Additionally, we note that the center of mass determined for each $n(x,\tau)$ shows interaction-dependent dynamics with similar features as $\delta n(x,\tau)$.

To improve the signal to noise ratio, we average $M$ measurements of $n(x,\tau)$ before determining the density deviation according to Eq.~\eqref{eq:density-dev}.
This procedure corresponds to the replacement $n(x, \tau) \rightarrow \sum_{i=1}^M n(x, \tau)^{(i)} / M$, where $n(x, \tau)^{(i)}$ labels an individual measurement.
In Fig~4(a), the number of measurements is constant across the shown $54$ points with \mbox{$M=4$} for all \mbox{$n(x, \tau\geq0)$}.
In Fig.~4(b), \mbox{$n(x,\tau=0)$} is determined from $6$-$15$~measurements, whereas \mbox{$n(x, \tau >0)$} is determined from $3$-$5$~measurements ($9$ points with \mbox{$M=4$}, $28$ points with \mbox{$M=4$}, and $17$ points with \mbox{$M=5$} measurements).

We estimate the uncertainty of $\delta n(\tau)$ with jackknife resampling.
To this end, we exclude one out of $M$ total measurements and determine $n(x, \tau)$ from the average of the remaining \mbox{$M-1$} measurements.
This yields $M$ distinct one-dimensional densities $n(x,\tau)$ from which we obtain the density deviation~$\delta n(\tau)$.
From the resulting distribution $\{\delta n(\tau)^{(i)}\}_{i=1,\ldots,M}$, we obtain an uncertainty of $\delta n$ from the jackknife estimate of the standard error~\cite{efron:1981}
\begin{align}\label{eq:jackknife-standard-error}
    \sigma_{\delta n}(\tau) = \sqrt{\frac{M-1}{M} \sum_{i=1}^M {\left[ \delta n(\tau)^{(i)} - \overline{\delta n(\tau)} \right]}^2},
\end{align}
where $\overline{\delta n(\tau)} = \sum_{i=1}^M \delta n(\tau)^{(i)} / M$.

Since the observable $\delta n(\tau)$ rectifies any differences between $n(x, 0)$ and $n(x, \tau)$, it is susceptible to technical noise.
To estimate the contribution of noise to our observable, we consider the density variance $\delta n^2$ and assume noise $n_\epsilon$ with vanishing mean $\langle n_\epsilon \rangle = 0$, finite variance $\langle n_\epsilon^2 \rangle > 0$, and vanishing third moment $\langle n_\epsilon^3 \rangle = 0$.
Here, $\langle X \rangle$ denotes the expectation value of the quantity~$X$.
From these assumptions and Eq.~\eqref{eq:density-dev}, we find that the contribution of $n_\epsilon$ to $\delta n^2$ is proportional to $\langle n_\epsilon^2 \rangle$.
We therefore expect for uncorrelated $n_\epsilon$,
\begin{align}\label{eq:dens-var-noise}
  \delta n_M^2(\tau) = \frac{\langle n_\epsilon^2 \rangle}{M} + \delta n^2(\tau) = \frac{\gamma}{M} + \delta n^2(\tau).
\end{align}
Here, $\gamma$ is a constant, $M$ is the number of individual measurements averaged for the calculation of \mbox{${n(x, \tau>0)}^2$} and $\delta n_M^2$ is the experimental measurement as opposed to the measurement-independent observable $\delta n^2$.
The value $\delta n_M^2$ for $M \leq 5$ can be estimated from subsampling since we take between three and five individual measurements for each parameter.
This allows us to extract the otherwise unknown value of $\delta n^2$ and $\gamma$ from a fit.
To avoid systematic errors, we use non-interacting data points in the stationary regime and find the noise contribution $\delta n_{M=3\ldots5}(\tau) - \delta n(\tau) \approx 0.4 \times 10^{-3}$, which is used to indicate the band of the theory data in Fig.~4.
We emphasize that this approach only serves as an estimate since it neglects potential correlations of~$n_\epsilon$.

\begin{figure}[t]
  \includegraphics[width=\columnwidth]{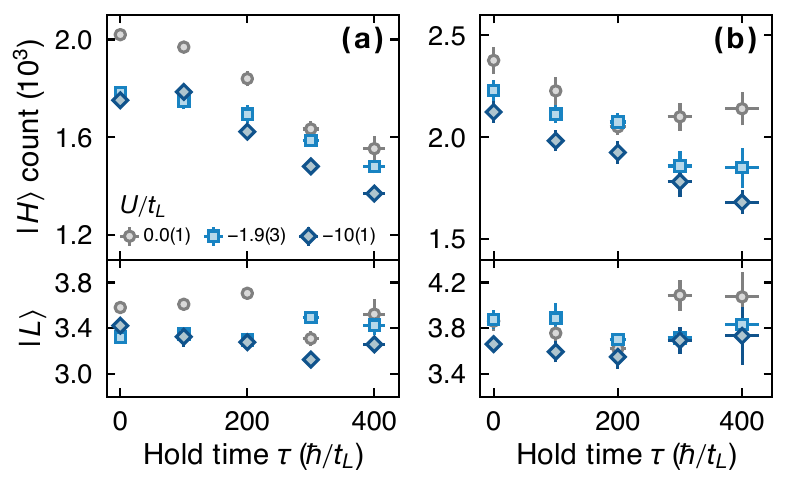}
  \caption{\label{fig:density-dev-loss}%
    Atom loss during the measurement of the density deviation~$\delta n(\tau, x)$.
    Number of heavy ($\ket{H}$, top panels) and light ($\ket{L}$, bottom panels) atoms during the measurement of the density deviation for (a)~$t_H/t_L =0.102(6)$ and (b)~$0.34(3)$ presented in Fig.~4.
    Data points spaced less than $50\hbar/t_L$ are binned to reduce visual clutter, and the interaction parameters are shown in the legend.
  }
\end{figure}

In Figs.~\subfigref{density-dev-loss}{c} and~\subfigref{density-dev-loss}{d}, we show the number of heavy and light atoms during the measurement of the density dynamics.
The initial atom numbers and loss rates of the heavy species are generally quite comparable for the different interaction parameters.

As discussed in the main text, we fit the experimental data shown in Fig.~4 with an exponential saturation curve, given by
\begin{align}\label{eq:exp-sat-fit}
  \delta n (\tau) = a \left( 1 - e^{-\tau/\tau_0} \right) + c.
\end{align}
Here, $a$, $c$ are dimensionless fit parameters, and $\tau_0$ is the fitted exponential decay constant, which is discussed in the main text.
By excluding each point individually in a separate fit routine [jackknife resampling, see Eq.~\eqref{eq:jackknife-standard-error}], we generate a distribution of $\tau_0$ from which we determine the estimated uncertainty of $\tau_0$ given in the main text.

To compare the two cases for $t_H/t_L\simeq0.1$ and finite interaction strength, we fit the measured density deviation to a linear function,
\begin{align}
    \delta n(\tau) = m \tau + c.
\end{align}
Restricting the numerical fit to early times, \mbox{$\tau\leq 101\hbar/t_L$}, we find the fitted parameters $m=13(2) \times 10^{-6}\, t_L / \hbar$, $2(2) \times 10^{-6}\, t_L / \hbar$ and $c=1.2(2)\times 10^{-3}\, t_L / \hbar$, $1.3(1)\times 10^{-3} \, t_L / t_H$ for the cases with $U/t_L \simeq -2$ and $-10$, respectively.
The same procedure yields $m\simeq 14 \times 10^{-6}\, t_L / \hbar$ and \mbox{$c\simeq 0.9 \times 10^{-3}\,t_L / \hbar$} for both finite interaction strengths with $t_H/t_L \simeq 0.34$.

\section{Theoretical description}\label{sec:theory}

\subsection{Non-interacting calculation}\label{sec:non-int-calc} 
For the theoretical calculations in the non-interacting limit (see Figs.~2 and~4), we employ the following single-particle Hamiltonian
\begin{align}\label{eq:single-particle-ham}
  \begin{split}
    \hat{\mathcal{H}}^{(1)} &= 
      -t_L \sum_i \left[ \hat{c}_{iL}^\dagger \hat{c}^{\phantom\dagger}_{(i+1)L}+ \text{h.c.} \right] \\ &\quad
      + \frac{\kappa}{2} \sum_i {\left(i-i_0\right)}^2 \hat{n}_{iL},
  \end{split}
\end{align}
which we numerically diagonalize for a finite but large system size $l = 250$ ($i \in [-l/2, l/2]$) to find the eigenenergies $\epsilon_k$ and eigenstates
\begin{align}
  \ket{\psi_k} = \sum_{i=-l/2}^{l/2} \alpha_i^{(k)} \hat{c}_{iL}^\dagger \ket{0}
\end{align}
with time evolution $\ket{\psi_k(\tau)} = e^{-\textrm{i} \epsilon_k \tau/\hbar} \ket{\psi_k}$.
To account for the finite temperature $T$ of the system (see Section~\ref{sec:lat-loading}), we calculate the occupation of the different eigenstates from the Fermi-Dirac distribution
\begin{align}
  f(\epsilon_k, T) = \frac{1}{e^{(\epsilon_k - \mu_0)/(k_B T)} + 1}.
\end{align}
Here, $T$ is the temperature, and the chemical potential $\mu_0 = \mu(N_L, T)$ is fixed by the constraint $\sum_k f(\epsilon_k, T) = N_L$ with $N_L$ the total number of light atoms in one tube.
Finally, we calculate the time-dependent density profile
\begin{align}
  n(i, \tau) = \sum_k f(\epsilon_k, T) {\left| \bra{\psi_k(\tau)} \hat{c}_{iL}^\dagger \ket{0} \right|}^2
\end{align}
from which we determine the experimentally relevant quantities.

The ramp of the trap minimum in the experiment is simulated by integrating the time-dependent Schr{\"o}dinger equation corresponding to the Hamiltonian~\eqref{eq:single-particle-ham} for all eigenstates.
For the long evolution times in Fig.~4, we average the calculation over ten equally spaced non-integer offsets~\mbox{$\in \left[-0.5, 0.5\right)$} of the trap minimum relative to $i_0$.
This is motivated by the fact that the relative phase of the lattice with respect to the minimum of the harmonic confinement is not fixed between individual shots of the experiment and also varies across the different tubes.
To approximately model the contributions of tubes with different atom numbers in the experiment, we use a binned distribution with different weights (see Section~\ref{sec:lat-loading}) calculated for the specific total number of light atoms $N_\text{tot}$ (see Table~\ref{tbl:exp-params}) and estimate our observable from a weighted average of these bins.
The theory curve in Fig.~2 systematically overestimates $N_r/N$, which can be corrected with a \mbox{$\approx 25\%$} increase of the estimated temperature or atom number.
We observe a similar deviation for the experimental and theoretical data in Fig.~4.
Overall, we attribute the disagreement to the approximative nature of our modeling of the lattice loading process.

\subsection{Many-body calculation}\label{sec:many-body-calc}
To study the interacting Hamiltonian defined in Eq.~(1) we employ tensor network methods as we aim to closely predict the dynamics observed in the experiment.
To account for the finite temperature of the system, we apply a density matrix-based approach, where the density matrix $\rho$ of the closed system is approximated as a matrix-product operator (MPO). 
By means of purification, this MPO can be expressed as pure matrix-product state (MPS) $\ket{\rho}$ in an enlarged Hilbert space, such that we obtain the density matrix by tracing out the ancilla degrees of freedom $\hat{\rho} = \Tr_a{\ket{\rho}\bra{\rho}}$. 
In this representation, we can make use of established tensor network algorithms developed in the MPS framework~\cite{verstraete:2004, zwolak:2004}. 
In particular, we implement our code based on the TenPy package~\cite{hauschild:2018}. 
To calculate the thermal equilibrium state $\hat{\rho} \propto e^{-\beta \hat{\mathcal{H}}}$ we start with the maximally mixed, infinite temperature state $\hat{\rho}_0 \propto \mathds{1}$, which can be represented exactly as a trivial MPO\@. 
For our nearest-neighbor Hamiltonian we employ the time evolving block decimation (TEBD) algorithm~\cite{paeckel:2019,vidal:2004} to obtain the equilibrium state at $T = 1/(k_B\beta)$,
\begin{align}
	\hat{\rho} \propto  \Tr_a{\left[e^{-\beta/2 \hat{\mathcal{H}}} \ket{\rho_0}\bra{\rho_0} e^{-\beta/2, \hat{\mathcal{H}}}\right]},
\end{align}
by evolving $\ket{\hat{\rho}_0}$ up to $\beta/2$ in imaginary time.
The dynamics of our system are governed by the von Neumann equation $\partial_\tau \hat{\rho} = -\mathrm{i} [\hat{\mathcal{H}}, \hat{\rho}]$, which we can write in terms of the purified MPS,
\begin{align}
	\partial_\tau \ket{\rho} = -\mathrm{i} \mathcal{L} \ket{\rho},
\end{align}
with the Liouville superoperator $\mathcal{L} = \hat{\mathcal{H}} \otimes \mathds{1} - \mathds{1} \otimes \hat{\mathcal{H}}$.
The solution $\ket{\rho(\tau)} = e^{-\mathrm{i} \mathcal{L} \tau} \ket{\rho}$ can be efficiently calculated by propagating the initial state in real time with the TEBD algorithm. 
The accuracy of our simulations depends critically on the operator-space entanglement, defined as the half-chain von Neumann entanglement entropy corresponding to the normalized MPS $\ket{\rho}$.
Throughout our simulations, we dynamically truncate the singular values at each bond, by keeping the discarded weight smaller than a threshold $\varepsilon_\mathrm{trunc}$, but keeping at most $\chi_\mathrm{max}$ Schmidt values. 
Note that in order to keep truncation errors negligible at fixed $\chi_\mathrm{max}$, our simulations are generally restricted to small atom numbers $N_L = N_H \leq 10$.

\subsubsection{Initial state}\label{sec:initial-state}
In general, we expect that the details of our initial state preparation in the experiment (see Section~\ref{sec:state-prep}) have an important influence on the dynamics.
Therefore, we study the preparation protocol numerically by approximating the time dependence of the Hubbard parameters [see Fig.~\ref{fig:exp-seq}] with a sequence of linear ramps and intermediate hold times.
We integrate the von Neumann equation for the time-dependent Hamiltonian by applying the TEBD algorithm with a first-order Trotter decomposition, while updating the time evolution operator within each simulation step.
This procedure resembles the simplest possible unitary integration scheme within our MPS approximation.
We start with a non-interacting thermal state at $T_\text{init} = 4t_L/k_B$ to approximately model the initial state in the \mbox{$\approx 7\Erec$} deep state-independent lattice, quench the hopping ratio $t_H/t_L = 1 \rightarrow 0.05$, apply the ramps for the confinement strength as well as hopping ratio, and subsequently turn on the interaction by ramping $U/t_L$ to the desired value.
\begin{figure}[t]
	\includegraphics[width=\columnwidth]{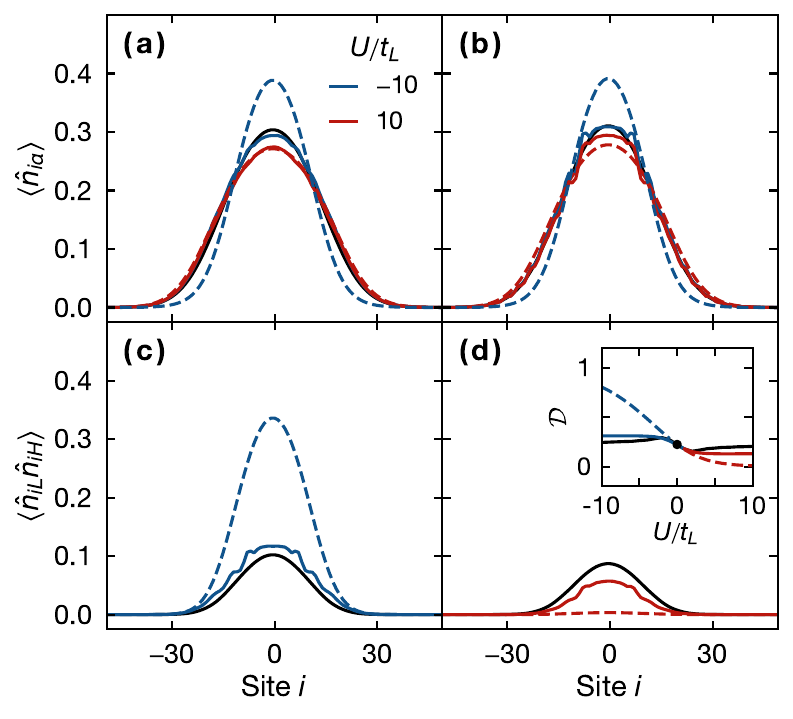}
	\caption{\label{fig:sim-state-prep}%
		Density profiles of (a)~light atoms ($\alpha=L$), (b)~heavy atoms ($\alpha=H$), and \mbox{(c),(d)}~doublons at the end of our state preparation protocol from a numerical simulation for a single tube with $N_L=N_H=10$ atoms.
    Solid colored lines correspond to the full state preparation protocol for $U/t_L = -10$ (attractive, blue) and $10$ (repulsive, red) with typical experimental parameters, while the dashed lines represent the corresponding thermal state at the $T_\text{eff}$ obtained from the local-density approximation and at finite interactions $U/t_L$.
    The black solid lines correspond to a non-interacting thermal state at  temperature $T_\text{eff}$.
    In the inset of panel~(d), we show the doublon fraction $\mathcal{D}$ for variable interaction strength $U/t_L$ and similar preparation protocols as in the main panels.
    Here, the solid black line corresponds to the non-interacting thermal state with an interaction quench $0 \rightarrow U/t_L$ and a subsequent hold time of~$10\hbar/t_L$.
	}
\end{figure}

In Fig.~\ref{fig:sim-state-prep}, we show the density profiles $\left\langle \hat{n}_{i\alpha} \right\rangle$ of the light atoms ($\alpha=L$) and heavy atoms ($\alpha=H$) as well as the density of doublons $\left\langle \hat{n}_{iL} \hat{n}_{iH} \right\rangle$ after calculating the state preparation for an exemplary $N_L=N_H=10$ tube with $U/t_L = -10$ and $10$.
Here, $\langle \hat{O} \rangle = \Tr(\hat{\rho} \hat{O})$ denotes the expectation value of the corresponding operator $\hat{O}$.
We compare the prepared state to a thermal equilibrium state at fixed temperature $T_\text{eff}$ for strong interactions and for the non-interacting case.
For this purpose, we choose the effective temperature of this state according to $T_\text{eff} = (\kappa_2 / \kappa_1) T_\text{init} \simeq 2.3 t_L/k_B$ where $\kappa_1$ ($\kappa_2$) is the trapping potential before (after) the state-independent optical lattice is ramped to the final depth (see Fig.~\ref{fig:exp-seq}).
As shown in Figs.~\subfigref{sim-state-prep}{a} and \subfigref{sim-state-prep}{b}, the density profiles of light and heavy atoms for the prepared state are similar to a thermal state at temperature $T_\textrm{eff}$ on the repulsive side, while significant differences become visible for the strong attractive case.
Furthermore, we put emphasis on the doublon density $\left\langle \hat{n}_{iL} \hat{n}_{iH} \right\rangle$ and relative fraction
\begin{align}
  \mathcal{D} = \frac{1}{N_{L}} \sum_{i} \left\langle \hat{n}_{iL} \hat{n}_{iH} \right\rangle\,.
\end{align}
Due to the expected exponentially long lifetime~\cite{sensarma:2010} and suppressed mobility of doublons~\cite{auerbach:1994} ($\tau_\mathcal{D} \propto t^2/U$, for the mass-balanced case $t=t_L=t_H$), we expect the initial fraction of doublons to influence the transport properties in our system critically.
Strikingly, the doublon fraction drastically deviates from the thermal state, which has a significantly higher (lower) value on the attractive (repulsive) side [see inset of Fig.~\subfigref{sim-state-prep}{d}].
Our calculations suggest that the system does not thermalize during the state preparation and the resulting state is still close to a non-interacting state at $T_\text{eff}$.
Moreover, the doublon fraction only weakly depends on the interaction strength, which is consistent with the symmetry between the attractive and repulsive side observed in the experiment [see Fig.~3].

\subsubsection{Simulation of the transport measurement}\label{sec:transport-sim}
Simulating the full dynamical response of our system even after displacing the trap minimum is generally limited by the rapidly growing operator space entanglement.
To numerically access long enough times before truncation errors become significant, we do not perform the full preparation protocol but rather choose the non-interacting thermal state at $T_\text{eff}$ as an approximate initial state and quench to the desired interaction strength.
Comparing the density profiles and the doublon fraction verifies that this approximates the prepared state reasonably well (see black lines in Fig.~\ref{fig:sim-state-prep}).
To further mitigate truncation errors due to growing entanglement, we only consider $N_L=N_H=5$ atoms in this simulation.
\begin{figure}[t]
	\includegraphics[width=\columnwidth]{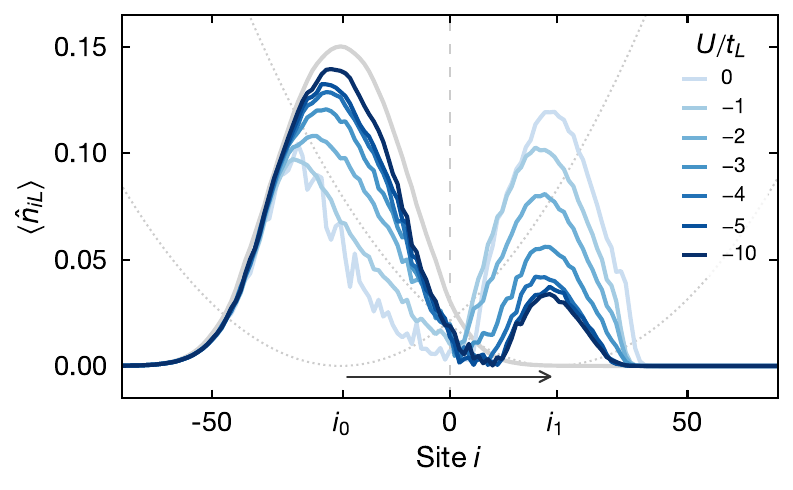}
	\caption{\label{fig:sim-transport}%
		Interaction-dependent transport of light atoms from a numerical simulation for a single tube of $N_L=N_H=5$ atoms, transport distance $45d$, and transport duration of $90\hbar/t_L$.
    The solid gray line corresponds to the initial state located at the initial trap minimum $i_0$ and blue lines represent the evolved density profiles for different interaction strength after translating the trap minimum to $i_1$.
    We indicate the initial and final harmonic confinement with dotted gray lines.
	}
\end{figure}
Fig.~\ref{fig:sim-transport} shows the transport after displacing the trap minimum over a distance of $45d$ within $90\hbar/t_L$ for different interaction strength on the attractive side.
The suppression of mobility with increasing $|U|/t_L$ becomes very clear and can be quantified by measuring the fraction of atoms on the right side of the system with $i \geq 0$, in analogy to the measurement of $N_r/N$ in the experiment.
The results for attractive and repulsive interactions are plotted in the inset of Fig.~3, and qualitatively agree with our experiment.

\subsection{Many-body Wannier-Stark localization}\label{sec:stark-mbl} 
As discussed in the main text, harmonic confinement in the experiment leads to Wannier-Stark localization of single-particle eigenstates with energies above the band edge at~$2t_L$.
Hence, transport in the non-interacting case is strongly suppressed, especially for the heavy atoms, as these experience a much stronger effective confinement $\kappa/t_H \gg \kappa/t_L$ compared to the light atoms.
For the interacting case, we expect some of these states to be delocalized through interactions~\cite{ott:2004, strohmaier:2007} already in the early time dynamics of our system.
Nevertheless, many-body Wannier-Stark localization~\cite{schulz:2019} is still expected for strong enough trapping and has been studied, e.g., in Ref.~\cite{chanda:2020} for the mass-balanced Fermi-Hubbard model (albeit relaxation at extremely late times cannot be fully ruled out).
Here, we extend these studies to the mass-imbalanced case.
To this end, we study the quench dynamics for an initial staggered product state
\begin{align}
	\ket{\Psi_0} = \ket{L~0~H~0~\dots} 
\end{align}
in a system of $64$ sites for a typical hopping ratio of $t_H/t_L=0.1$.
We express the initial state as an exact MPS and calculate the time evolution with the TEBD algorithm.
\begin{figure}[t]
	\includegraphics[width=\columnwidth]{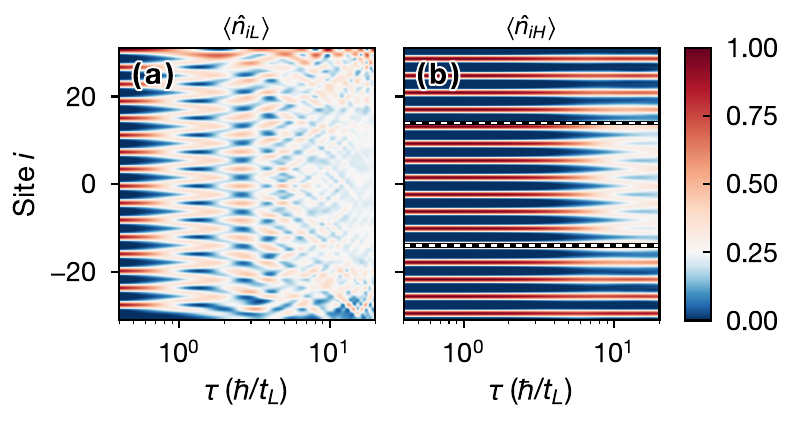}
	\caption{\label{fig:stark_mbl}%
		Time-dependent quench dynamics of the atomic densities for an initial staggered state in a quarter filled system of size $l=64$, shown for (a)~light and (b)~heavy atoms.
        The dashed lines in panel~(b) indicate the boundary between localized and delocalized states.
        The Hubbard parameters are close to typical experimental values with $\kappa/t_L = 2\times10^{-2}$, $t_H/t_L = 0.1$, and $U/t_L=-1.0$.
        We note that the visible asymmetry on the edges of the system originates from the non-inversion-symmetric initial state.
	}
\end{figure}

Fig.~\ref{fig:stark_mbl} depicts the dynamics of the atomic densities for both species. 
The light atoms equilibrate on a time scale $\hbar/t_L$, and we observe relaxation for the given system size and confinement $\kappa/t_L = 2\times10^{-2}$.
In contrast, relaxation of heavy atoms starts at a time scale $\hbar/t_H$ and occurs in the center of the trap.
Close to the edges of the system, the local gradient leads to a confinement of the heavy atoms on the simulated time scales.
For the mass-balanced Fermi-Hubbard model, the local critical potential has been estimated as $\partial_i \mathcal{H} / \hat{n}_{i\alpha} = \kappa\,(i_\mathrm{crit} - i_0) \simeq 2.8$~\cite{chanda:2020} which agrees well with our observations in the mass-imbalanced case.
We therefore estimate that already at short times the heavy atoms are mobile over a central trap region of $2i_\mathrm{crit} \approx 30$ lattice sites.
Our simulations explicitly focus on weak interactions $U/t_L=-1.0$ to ensure that dynamical constraints arising from doublon physics and mass imbalance are weak, and we consequently only probe the influence of harmonic confinement on the mobility of the heavy atoms.

\begin{figure}[t]
	\includegraphics[width=\columnwidth]{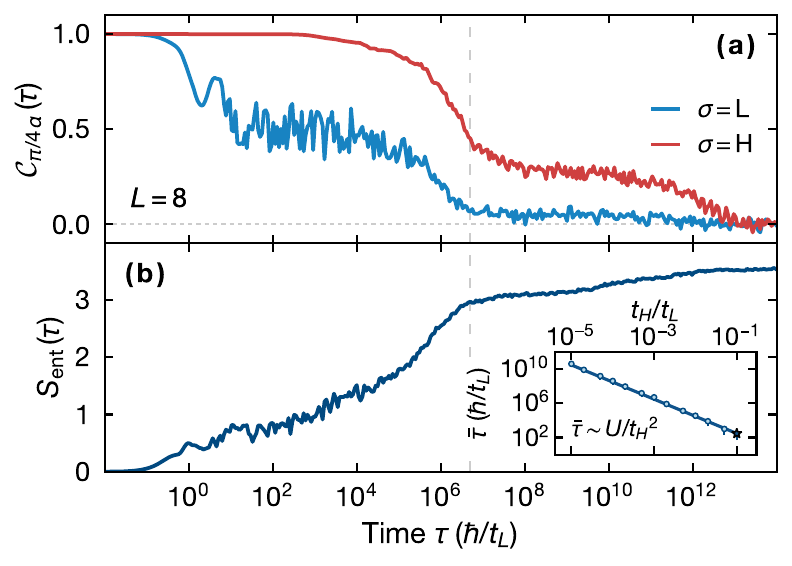}
	\caption{\label{fig:t3_estimate}%
	Extraction of time scales for large mass imbalance.
    (a)~Decay of a long-wavelength excitation and (b)~growth of entanglement entropy for initial product states evaluated for a translational-invariant, periodic system of size $l=8$ for $N_L=N_H=4$ atoms.
    We consider the interaction strength~${U/t_L=-10}$ for the strong mass-imbalanced case of ${t_H/t_L=10^{-3}}$ to illustrate the different time scales.
    The inset shows the time scale ${\bar\tau = C \times \hbar U / t_H^2}$ corresponding to the onset of a regime of unconventionally slow relaxation [dashed line in panels~(a) and~(b)].
    Here, the constant $C$ is extracted from a fit displayed as solid lines.
    We indicate the experimentally relevant value for~$t_H/t_L~=~0.1$ with a star-shaped marker.
	}
\end{figure}

\subsection{Time scales}
It has been discussed that strong mass imbalance in small translational-invariant ($\kappa=0$) systems leads to an ergodic regime of extremely slow equilibration~\cite{yao:2016}.
For the delocalized atoms in the center of the trap, we expect the dynamics to be close to this $\kappa \rightarrow 0$ limit.
This suggests that our finite-size experimental system with large mass imbalance will feature similar physics.
To study the involved time scales, we turn to very large mass imbalance and long evolution times.
For this purpose, we perform exact diagonalization of the translational-invariant model with the Hamiltonian in Eq.~(1) taking the form,
\begin{align}\label{eq:hamiltonian}
	\hat{\mathcal{H}^\prime} = -\sum_{i,\alpha\in\{L,H\}} t_\alpha \left[\hat{c}_{i\alpha}^\dagger \hat{c}^{}_{(i+1)\alpha} + \text{h.c.}\right] + U \sum_i \hat{n}_{iL} \hat{n}_{iH},
\end{align}
subject to periodic boundary conditions, $\hat{c}^\dagger_{l\alpha} = \hat{c}^\dagger_{0\alpha}$, for a system of size $l=8$ at half filling $N_L=N_H=4$.
To estimate the relaxation time of a density modulation, we calculate the dynamical correlator
\begin{align}\label{eq:dyn-corr}
	\mathcal{C}_{q\alpha}(\tau) = \left\langle \hat{n}_{q\alpha}^\dagger \hat{n}_{q\alpha}^{\vphantom{\dagger}}(\tau) \right\rangle,
\end{align}
which measures the decay of an inhomogeneous density perturbation $\hat{n}_{q\alpha} = \sum_{j} e^{-\mathrm{i} q j} \hat{n}_{j\alpha} $ for one species (${\alpha = L,H}$).

The decay of long-wavelength modulations is shown in Fig.~\subfigref{t3_estimate}{a} for $t_H/t_L=10^{-3}$ and indicates the existence of distinct time scales governing the dynamics.
These time scales also become apparent in the entanglement dynamics corresponding to plateau-like features of very slow entanglement growth [see Fig.~\subfigref{t3_estimate}{b}].
Here, we compute the half-chain entanglement entropy for random initial product states, averaging the results of $100$ states.
After initial relaxation of the light particles on the time scale~\mbox{$\sim \hbar/t_L$} the dynamics stagnate, and the entanglement entropy stays nearly constant up to time~\mbox{$\sim \hbar/t_H$}.
Following further relaxation and an increase of the entanglement entropy, the system enters a new regime at time scale~\mbox{$\sim \hbar U/t_H^2$} [dashed lines in Figs.~\subfigref{t3_estimate}{a} and~\subfigref{t3_estimate}{b}].
This regime features slow decay of non-equilibrium states~\cite{yao:2016}, and we observe an accordingly slow growth of entanglement entropy.
Although the phenomenology shows similarities to a dynamical type of many-body localization, the system is still ergodic, and density perturbations eventually decay.
From our exact diagonalization data, we extract the time scale $\bar\tau = C \times  \hbar |U|/t_H^2$ for the onset of this regime and estimate $C = 0.29(3)$ from a fit [see Fig.~\subfigref{t3_estimate}{c}].

\begin{figure}[t]
	\includegraphics[width=\columnwidth]{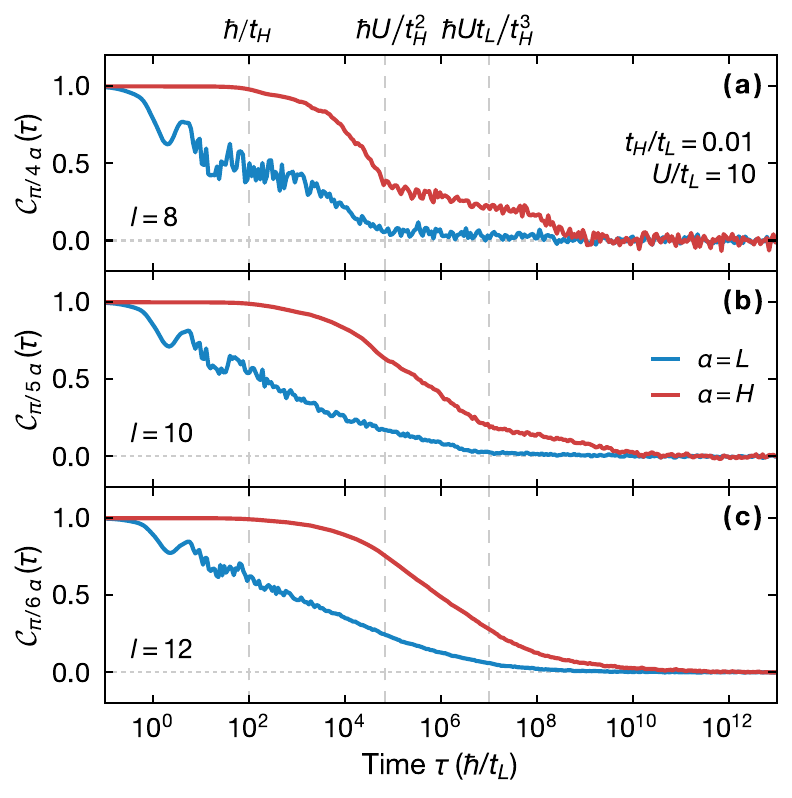}
	\caption{\label{fig:ed-finite-size-scaling}%
    System-size dependence of the dynamical correlator $\mathcal{C}_{q\alpha}(\tau)$ for translational-invariant and periodic systems of size (a)~$l=8$, (b)~$l=10$, and (c)~$l=12$.
    The wavevector~$q$ correspond to the longest wavelength $\lambda=2\pi/q$ supported by the respective system size under periodic boundary conditions.
    Here, the strongly interacting case at $U/t_L=-10$ with the mass ratio $t_H/t_L = 10^{-2}$ is considered.
    Note that the features of the decay dynamics become less pronounced with increasing system size.}
\end{figure}
%
\begin{figure*}[t]
	\includegraphics[width=\textwidth]{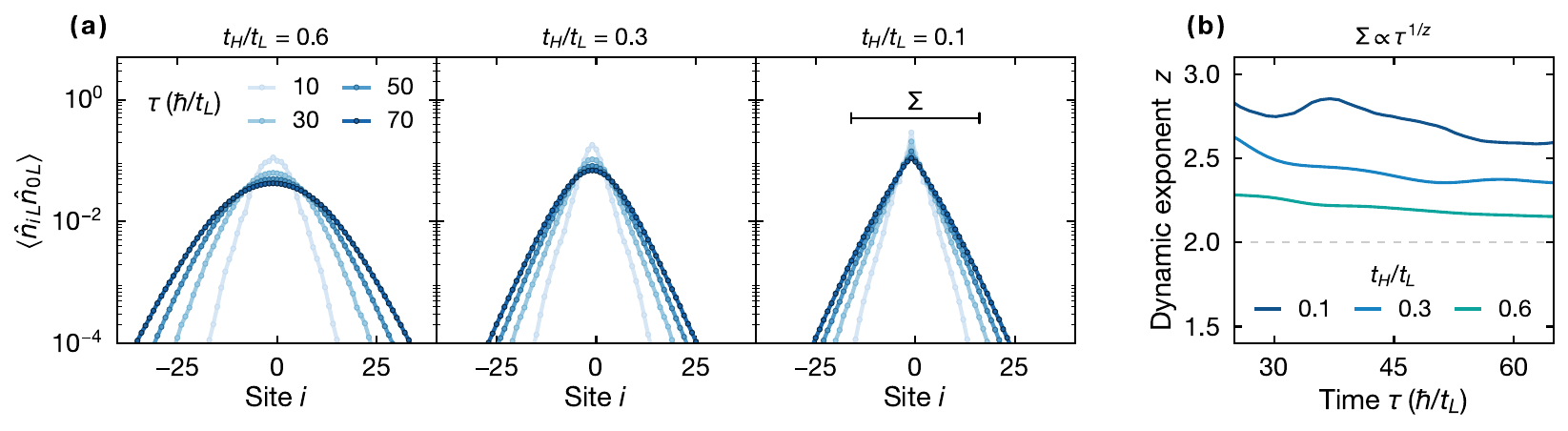}
	\caption{\label{fig:anormalous-transport}%
    Transient transport properties of the mass-imbalanced Fermi-Hubbard model.
    (a)~Spatial profiles of the density-density correlation function of the light species for different mass-ratios ${t_H/t_L=0.6, 0.3, 0.1}$ (left to right) at four points in time.
    (b)~Scaling of the spatial width $\Sigma_{L}(\tau) \propto \tau^{1/z}$ obtained from the mean squared displacement.
    The dynamic exponent $z$ is obtained by power law fits to the time dependent spatial width $\Sigma_{L}(\tau)$ over time windows of size $\Delta \tau = 10/t_L$.
    For ordinary diffusion Gaussian, profiles and dynamical exponents $z=2$ are expected, while $z>2$ indicates subdiffusion.
	}
\end{figure*}
We note that for increasing $t_H/t_L$ and increasing system size, the features separating different regimes become less pronounced.
In Fig.~\ref{fig:ed-finite-size-scaling} we illustrate the finite-size scaling for systems of size $l=8$, $10$, and $12$.
When the system size is increased new time scales enter the dynamics and additional features in the relaxation dynamics and entanglement growth become visible.
This is exemplified for~${l=10}$ in Fig.~\subfigref{ed-finite-size-scaling}{b}, where the time scale~$\sim\hbar U t_L/t_H^3$ is observed.
Such a hierarchy of time scales ``washes'' out distinct plateau-like features, which for large systems leads to featureless decay dynamics.
Note that distinct time scales in the relaxation dynamics are only expected for small systems or extremely small mass ratios~$t_H/t_L$~\cite{papic:2015}.

Few-body bound states constitute a promising explanation for the observed time scales in small systems.
A bound state of $N_H$~heavy and $N_L$~light atoms is stable for~${t_H/t_L<1}$, and can become mobile at time~${\sim \hbar \, U^{N_L} t_L^{N_H-N_L-1} / t_H^{N_H}}$, bottlenecking the overall dynamics.
The emergence of such bound states could explain the unconventionally slow relaxation, where a cascade of time scales leads to the observed dynamics.

\subsection{Properties of the transient regime}\label{sec:transient-regime}
In our experimentally realized Fermi-Hubbard model, mass imbalance between the two species induces strong dynamical constraints on the light atoms.
Contrary to the small size exact diagonalization studies discussed in the previous section, the experiment is conducted at moderate mass-ratios ${t_H/t_L \geq 0.05}$, and in a significantly larger system than accessible to exact diagonalization.
In this case, we expect overall slow relaxation dynamics, but no distinct time scales in form of plateau-like features.

Here, we briefly discuss the transport properties of our system in this transient regime of unconventionally slow relaxation.
To this end, we consider the Hamiltonian in Eq.~(1) for mass ratios $t_H/t_L \geq 0.1$ with vanishing confinement $\kappa/t_L = 0$, and moderate attractive interactions $U/t_L = -4$.
Within the accessible simulation time, we find the features to be most pronounced for this intermediate interaction strength, while other values of $U/t_L$ lead to qualitative similar behavior.
To characterize transport properties, we calculate the density-density correlation function
\begin{align}\label{eq:dyn-corr}
	\mathcal{C}_{i\alpha}(\tau) = \left\langle \hat{n}_{i\alpha} \hat{n}_{0\alpha}(\tau) \right\rangle.
\end{align}
We simulate the Heisenberg time evolution of the operator $\hat{n}_{0\alpha}$ with the same method as discussed for the density matrix (see Section~\ref{sec:many-body-calc}).
From $\hat{n}_{0\alpha}(\tau)$ the full spatial correlation profile $\mathcal{C}_{i\alpha}(\tau)$ can be obtained due to translational invariance.
We note, that for the Heisenberg equation of motion $\partial_\tau \hat{O} = \mathrm{i}[\hat{\mathcal{H}}, \hat{O}]$, the superoperator reads $\mathcal{L} = -\hat{\mathcal{H}} \otimes \mathds{1} + \mathds{1} \otimes \hat{\mathcal{H}}$.
The simulations are carried out for a system of size $l=201$, 
which has been verified not to exhibit finite-size effects on the accessible simulation times.

Transport properties can be characterized by the scaling of the mean squared displacement
\begin{align}\label{eq:msd}
  \Sigma^2_\alpha(\tau) = \sum_i i^2 \, \mathcal{C}_{i\alpha}(\tau) - {\left[\sum_i i\,\mathcal{C}_{i\alpha}(\tau) \right]}^2 ,
\end{align}
which quantifies the spatial width $\Sigma_\alpha(\tau)$ of the correlation profile.
For diffusion, we expect the spatial profiles to be well described by Gaussians~\cite{bertini:2021}, with the spatial width scaling as ${\Sigma_\alpha(\tau) \propto \tau^{1/2}}$.
Generally, a power law scaling ${\Sigma_\alpha(\tau) \propto \tau^{1/z}}$ indicates ballistic and diffusive transport, for $z=1$ and $z=2$, respectively,
superdiffusion for $1<z<2$, and subdiffusion for $z>2$~\cite{bertini:2021}.

In Fig.~\subfigref{anormalous-transport}{a}, spatial profiles of the density-density correlator corresponding to the light species are shown for different mass ratios and at different times.
For small mass ratios we find that, within the accessible simulation time, the spatial profiles cannot be described by Gaussians, but rather have a peaked shape in the semi-logarithmic plots.
This observation indicates anomalous transport in the system, and has previously been found in small systems with exact diagonalization~\cite{heitmann:2020}.

Moreover, the broadening of the spatial width with time slows down with increasing mass imbalance.
To characterize this regime, we extract the dynamic exponent $z$ from the logarithmic derivative of $\Sigma_{L}(\tau)$ averaged over time windows of fixed size $\Delta \tau = 10$.
As shown in Fig.~\subfigref{anormalous-transport}{b}, this reveals subdiffusive behavior $z>2$ of our system on the accessible simulation times.
However, we observe that the dynamic exponent $z$ is time-dependent and decreases with time.
The value of $z$ approaches $2$ at late times, which indicates a crossover from subdiffusive ($z>2$) to diffusive dynamics ($z=2$).
These findings suggest that our experimental system exhibits a transient regime of slow subdiffusive relaxation dynamics, as indicated in the suggested dynamical phase diagram depicted in Fig.~1(c).

\bibliography{references}